%% file: 2026-Frank-RegionallyStableRnn.tex
\documentclass{ifacconf}

\usepackage{graphicx}      
\usepackage{natbib}        
\usepackage{enumitem}

\input{utils/preamble}

\input{utils/glossary}
\input{utils/tikz}

\newif\ifpng
\pngfalse

\newcommand{\DF}[1]{#1}

\begin{document}

\begin{frontmatter}

\title{
Learning the dynamics of nonlinear systems with regional stability guarantees through linear matrix inequality constraints\thanksref{footnoteinfo}} 

\thanks[footnoteinfo]{Daniel Frank is funded by Deutsche Forschungsgemeinschaft (DFG, German Research Foundation) under Germany's Excellence Strategy - EXC 2075 - 390740016. We acknowledge the support of the Stuttgart Center for Simulation Science (SimTech). © 2026 the authors. This work has been accepted to IFAC for publication under a Creative Commons Licence CC-BY-NC-ND.}

\author[First]{Daniel Frank} 
\author[Second]{Fahim Shakib} 
\author[First,Third]{Steffen Staab}

\address[First]{Institute for Artificial Intelligence, University of Stuttgart, Germany (e-mail: \{daniel.frank, steffen.staab\}@ki.uni-stuttgart.de}
\address[Second]{Department of Mechanical Engineering, Eindhoven University of Technology, Eindhoven, The Netherlands (e-mail: m.f.shakib@tue.nl)}
\address[Third]{Electronics and Computer Science, University of Southampton, United Kingdom of Great Britain and Northern Ireland}

\begin{abstract}                
This paper presents a method that learns a regionally stable recurrent neural network model from a set of input-output data generated by an unknown dynamical system.
Relying on \emph{generalized sector} conditions on the deadzone activation function, we first derive sufficient conditions that guarantee forward invariance on a compact set of the state space for any inputs from a given set.
Such regional properties lead to less conservative conditions compared to variants that offer a global form of stability, and are in line with the system data that is only observed regionally.
Our learning method derives
conditions for regional stability using a barrier function approach, leading to models equipped with a certificate of regional stability in a subset of the state space and for a given input set.
We illustrate our theoretical result with a numerical example and compare it to methods that impose a global form of stability, which fail to identify the system, and with a method that imposes no stability constraints at all, which does not guarantee a stable behavior within any state or input set.
\end{abstract}


\end{frontmatter}

\section{Introduction}
\label{sec:intro}
\input{content/introduction}

\section{Problem Statement}
\label{sec:problem}

\input{content/problem}

\section{Regional stability analysis for nonlinear system identification}
\label{sec:analysis}

\input{content/analysis}

\section{Training}
\label{sec:training}

\input{content/training}

\section{Numerical case study}
\label{sec:evaluation}

\input{content/evaluation}

\section{Conclusion}
\label{sec:conclusion}
\input{content/conclusion}

\begin{ack}
The author thanks Dr.~Tobias Holicki, who gave the initial ideas and has helped to develop the analysis part.
\end{ack}

\section*{DECLARATION OF GENERATIVE AI AND AI-ASSISTED TECHNOLOGIES IN THE WRITING PROCESS}
During the preparation of this work, \emph{Claude Sonnet 4.5 in Visual Studio Code} was used for coding assistance. After using this tool, the author(s) reviewed and edited the content as needed and takes full responsibility for the content of the publication.

\bibliography{ifacconf}             
                                                   







\appendix

\end{document}

%% file: utils/preamble.tex
\usepackage{url}            
\usepackage{booktabs}       
\usepackage{amsfonts}       
\usepackage{enumitem}

\usepackage{amsmath,amssymb}
\usepackage[acronym]{glossaries} 
\usepackage{tikz} 
\usepackage{mathtools} 
\usetikzlibrary{matrix, shapes, arrows, positioning, chains, patterns, fit, decorations.pathreplacing, calc, plotmarks, arrows.meta}
\usepackage{pgfplots} 
\pgfplotsset{compat=1.18}
\usepackage{subcaption}
\DeclareCaptionLabelFormat{subsimple}{#2#1}
\usepackage{arydshln}
\usepackage{algpseudocode}
\usepackage{algorithm}
\usepgfplotslibrary{groupplots,dateplot}

\newcommand{\mat}[2]{\left(\begin{array}{@{}#1@{}}#2\end{array}\right)}
\newcommand{\smat}[1]{\left(\begin{smallmatrix}#1\end{smallmatrix}\right)}
 
\newcommand{\mstrut}[1]{\rule{0pt}{#1}}
\let\oldhline=\hline 
\renewcommand{\hline}{\oldhline\mstrut{2.5ex}}
\let\oldhdashline=\hdashline 
\renewcommand{\hdashline}{\oldhdashline\mstrut{2.5ex}}
 
\newcommand{\FS}[1]{#1}

\newcommand{\T}{\mathrm{T}}

\newcommand{\Dc}{\mathcal{D}}

\newcommand{\Xc}{\mathcal{X}}

\newcommand{\Sc}{\mathcal{S}}
\newcommand{\Rc}{\mathcal{R}}

\newcommand{\Lc}{\mathcal{L}}
\newcommand{\Ec}{\mathcal{E}}
\newcommand{\Uc}{\mathcal{U}}

\newcommand{\Cb}{\mathbf{C}}

\newcommand{\Fb}{\mathbf{F}}
\newcommand{\Gb}{\mathbf{G}}
\newcommand{\Lb}{\mathbf{L}}

\newcommand{\Rbb}{\mathbb{R}}
\newcommand{\Dbb}{\mathbb{D}}
\newcommand{\Sbb}{\mathbb{S}}
\newcommand{\Nbb}{\mathbb{N}}

\newcommand{\dzn}{\operatorname{dzn}}
\newcommand{\sat}{\operatorname{sat}}

\newcommand{\MGenSec}{\textsc{GenSec} (ours)}
\newcommand{\MStdSec}{\textsc{StdSec}}
\newcommand{\MLtiRnn}{\textsc{NoSec}}

\newcommand{\nx}{n}
\newcommand{\nz}{m}
\newcommand{\nd}{r}
\newcommand{\ny}{e}
\newcommand{\nw}{q}

\newcommand{\x}{x}
\newcommand{\dd}{u}
\newcommand{\w}{w}

\newcommand{\ee}{y}
\newcommand{\z}{v}

\newcommand{\xk}[1]{x_{#1}}
\newcommand{\dk}[1]{u_{#1}}
\newcommand{\wk}[1]{w_{#1}}

\newcommand{\ek}[1]{y_{#1}}
\newcommand{\eHatK}[1]{\hat{y}_{#1}}
\newcommand{\zk}[1]{v_{#1}}



%% file: utils/glossary.tex
\newacronym{rnn}{RNN}{recurrent neural network}
\newacronym{lstm}{LSTM}{long-short term memory}
\newacronym{relinet}{ReLiNet}{Recurrent Linear Parameter Varying Network}
\newacronym{arx}{ARX}{autoregressive with exogenous inputs}
\newacronym{tuv}{TÜV}{technical inspection association}
\newacronym{lti}{LTI}{linear, time-invariant}
\newacronym{l4dc}{L4DC}{Learning for Dynamics \& Control Conference}
\newacronym{ijcai}{IJCAI}{International Joint Conferences on Artificial Intelligence}
\newacronym{iss}{ISS}{input-to-state stability}
\newacronym
  [longplural={linear matrix inequalities}]
  {lmi}{LMI}{linear matrix inequality}
\newacronym{lft}{LFT}{lower linear fractional transformation}
\newacronym{lfr}{LFR}{Linear Fractional Representation}
\newacronym{ood}{OOD}{out-of-distribution}
\newacronym{id}{ID}{in-distribution}
\newacronym{sdp}{SDP}{semidefinite program}
\newacronym{dof}{DOF}{degrees-of-freedom}
\newacronym{msd}{MSD}{mass-spring-damper}
\newacronym{roa}{RoA}{Region of Attraction}
\newacronym{lqr}{LQR}{linear quadratic regulator}
\newacronym{rbf}{RBF}{radial basis function}
\newacronym{grf}{GRF}{Gaussian random field}
\newacronym{siso}{SISO}{single-input-single-output}
\newacronym{mimo}{MIMO}{multiple-input-multiple-output}
\newacronym{mse}{MSE}{mean-squared-error}

%% file: utils/tikz.tex
\tikzset{
    mass/.style={
        draw,
        thick,
        minimum width=3em,
        minimum height=3em
    },
    spring/.style={
        thick,
        decorate,
        decoration={
            zigzag,
            pre length=0.3cm,
            post length=0.3cm,
            segment length=6
        }
    },
    ground/.style={
        fill,
        pattern=north east lines,
        draw=none,
        minimum width=0.75cm,
        minimum height=0.3cm
    },
    dampic/.pic={
        \fill[white] (-0.1,-0.2) rectangle (0.2,0.2);
        \draw (-0.2,0.2) -| (0.2,-0.2) -- (-0.2,-0.2);
        \draw[line width=1mm] (-0.1,-0.2) -- (-0.1,0.2);
    },
    dotted_block/.style={
        draw=black!30!white, 
        dashed,
        inner ysep=0.6em,
        inner xsep=2.1em, 
        rectangle, 
        rounded corners
    },
    block/.style={
        draw,
        rectangle,
        rounded corners,
        minimum height=2em,
        minimum width=2em,
        inner ysep=0.5em,
        inner xsep=0.5em
    },
    sum/.style={
        draw,
        circle, 
        node distance=1.5cm
    },
    operator/.style={
        draw,
        circle,
        thin,
        minimum height=1em,
	   inner sep=1pt
    },
    weight/.style={
        draw,
        thin,
        rounded corners,
        rectangle,
    },
    value/.style={
        draw,
        thin,
        rectangle,
    },
    gain/.style={
        regular polygon, 
        regular polygon sides=3,
        draw, 
        fill=white, 
        text width=1em,
        inner sep=1mm, 
        outer sep=0mm,
        shape border rotate=-90
    },
    concat/.style={
        draw,
        shape=circle, 
        fill=black,
	   inner sep=0pt
    },
}

%% file: content/introduction.tex
System identification finds
a dynamic model from input and output data by minimizing some loss function, e.g., the \gls{mse} between predicted outputs and the outputs in the dataset~\citep{ljung1999system}.
\Glspl{rnn} are particularly
well-suited for learning complex nonlinear dynamics
\glspl{rnn}~\citep{beintema2023deep, mohajerin2019multistep, bonassi2022recurrent, revay2020convex}.
Although accurate on the training data, \Glspl{rnn} can fail to generalize to unseen test data, even leading to unstable prediction, which is undesired especially in safety-critical applications.
Consequently, there is a shift towards learning with \emph{stability guarantees} \citep{pauli2021training, revay2023recurrent,shakib2022computationally}.

Existing methods typically enforce a form of \emph{global} stability.
For example,~\cite{revay2020convex, revay2023recurrent} and \cite{yin2021stability} enforce global contraction,~\cite{shakib2025numerical} enforces global convergence, while~\cite{bonassi2022recurrent} reviews methods that enforce a global form of (incremental) input-to-state stability.
Enforcing a global form of stability
relates to using the \emph{standard} sector condition, see~\citep{Khalil:1173048}, to bound the effect of the nonlinearity globally.

In practice, enforcing a global notion of stability has two main drawbacks.
First, measured system data is only observed in a region of the input and state space, making it impossible to verify the stability properties of the data-generating system globally.
Secondly, conditions for global stability are overly conservative.
To address these issues, adopting a regional form of stability is beneficial: it allows for preserving the observed stability property more accurately and enlarges the set of learnable models, since regional stability constraints are less restrictive than global ones~\citep{shakib2023kernel}.
Despite these benefits, learning nonlinear dynamics with a regional form of stability is an open problem.
Notably, the work in~\cite{shakib2023kernel} proposes a kernel-based method that enforces regional convergence, a property ensuring forward invariance of a compact set for a given set of inputs and guaranteeing that all trajectories originating within this set converge to each other exponentially.

In this paper, we propose a learning method that enforces regional stability of the RNN.
In particular, we employ \emph{generalized} sector conditions, which introduce a degree of freedom to fit the sector to the deadzone nonlinearity tightly and are well-established in anti-windup control (see~\cite{tarbouriech2006stability}) as well as in the control design with RNNs~\citep{la2025regional}.
Notably, in these related works, the models are autonomous, i.e., there is no external excitation.
Therefore, their direct use in the scope of learning system with inputs is infeasible and requires a non-trivial extension.
To address this gap, our first contribution is a convex constraint, in the form of \glspl{lmi}, on the learnable parameters that ensures forward invariance of a compact set of the state space for a given set of inputs.
Furthermore, these conditions provide a regional input-to-state stability bound for all trajectories originating within this set.

The second contribution is a learning framework that enforces the derived regional-stability conditions on the learned RNN. 
\DF{Hereto, we equip the \gls{mse} loss with a barrier function to enforce the convex constraints during learning.}
This approach is scalable to models with a large number of parameters, as no \glspl{lmi} need to be solved during training.
The resulting RNN is regionally stable, with the most suitable compact state space set, i.e., the region, emerging from the training process for a given input set.
This learning method requires an initial RNN that already satisfies the convex constraints, for which we present a separate result. 

The contribution of our work can be summarized as follows:
\begin{itemize}
    \item We provide convex constraints for regional stability of RNNs with external inputs.
    \item We provide a learning framework that enforces the constraints using a barrier function approach.
\end{itemize}
The theoretical results are supported by a numerical experiment, which demonstrates recovering a regionally stable system from input-output data. 
We compare our approach with a method that enforces a global form of stability, which fails to identify the system, and a method that does not enforce any form of stability, which lacks theoretical guarantees. All proofs are omitted due to space limitations.

\textbf{Notation:}
We define the sets of real and natural numbers as $\Rbb$ and $\Nbb$, respectively, with $\Nbb_0 = \Nbb \cup \{0\}$.
For $\x \in \mathbb{R}^n$, we define $|x|^2 \coloneqq x^\top x$ and $\| x \|_\infty = \sup_i |x^i|$, where $x^i$ is the $i$-th element of $x$.
We use $\star$ to denote symmetric terms in matrices. With $\Dbb_+$, we denote the set of positive-definite diagonal matrices and $\Sbb_+$ the set of positive-definite symmetric matrices. 
Given a positive-semidefinite matrix $X$, we define the ellipsoid $\Ec(X) \coloneqq \{x\in\Rbb^n | x^\T X x \leq 1\}$.
We consider sequences in the extended $\ell_{2e}^n$ space and define it as
$\ell_{2e}^n\coloneqq \{(\xk{k})_{k\in\Nbb_0} \mid \xk{k}\in \Rbb^n \mathrm{\ and\ } \sum_{k=0}^N |\xk{k}|^2 < \infty \mathrm{\ for\ all\ } N\in\Nbb_0\}$.
Furthermore, for any real-valued matrix $H\in \mathbb{R}^{p\times n}$, we define the set $\Lc(H) \coloneqq \{x\in\Rbb^n \,\vert\, \|H x\|_\infty \leq 1\}$.
 We use $X\sim\Rc(a,b)$ to denote that the random variable $X$ is drawn from a uniform distribution over the interval $(a,b)\subset \Rc$.

%% file: content/problem.tex
We are given a set of $M$ input-output trajectories from an unknown data-generating system, including an initial state $\xk{0}$ defined as follows:
\begin{equation}
    \label{eq:dataset}
    \Dc = \left\{ \left( \left( \ek{k}, \dk{k}\right) _{k=0}^{N_i-1} , \xk{0} \right)^{[i]} \right\}_{i=1}^M,
\end{equation}
where $\left(\left( \ek{k}, \dk{k}\right)_{k=0}^{N_i-1} , \xk{0} \right)$ is a single trajectory of length $N_i$ and $k\geq 0$ denotes the discrete time index. The trajectories $\dd$ and $\ee$ refer to the input and output respectively with $\dk{k} \in \Rbb^{\nd}$, $\ek{k}\in \Rbb^{\ny}$, and $\xk{0} \in \Rbb^{\nx}$.

\begin{figure}
    \centering
    \input{fig/model}
    \caption{Model~\eqref{eq:state-space-model} with static nonlinear function $\Delta$.}
    \label{fig:nonlinear-state-space-model}
\end{figure}
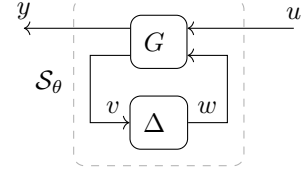

With the dataset $\Dc$, we aim to identify a nonlinear state space model $\Sc_\theta$ of the form
\begin{subequations}
\label{eq:state-space-model}
\DF{
\begin{equation}
    \label{eq:linear-system}
    G :=\left\{
    \mat{c}{\xk{k+1}\\ \eHatK{k} \\ \zk{k}} = 
    \mat{ccc}{A & B & B_{2}\\C & D & D_{12} \\ C_{2} & D_{21} & 0}
    \mat{c}{\xk{k}\\\dk{k} \\ \wk{k}}  \right.
\end{equation}}
with a static nonlinear function
\begin{equation}
    \label{eq:nonlinearity}
    \wk{k} = \Delta(\zk{k}),
\end{equation}
\end{subequations}
where $\xk{k} \in \Rbb^{\nx}$ refers to the internal state at discrete time step $k$. The external input and predicted output are denoted by $\dk{k}\in \Rbb^{\nd}$ and $\eHatK{k} \in \Rbb^{\ny}$, respectively. 
The output $\zk{k}\in \Rbb^{\nz}$ is fed through the nonlinear function $\Delta$ and enters~\eqref{eq:linear-system} as $\wk{k}\in\Rbb^{\nw}$. 
The nonlinear activation function $\Delta: \Rbb^{\nz} \rightarrow \Rbb^{\nw}$ applies the same scalar function $\psi:\Rbb\rightarrow \Rbb$ elementwise; hence, $\nw = \nz$.
A block diagram of model \eqref{eq:state-space-model} is shown in Figure \ref{fig:nonlinear-state-space-model}.
The model parameters are collected in $\theta = \{A,~B,~B_2,~C,~D,~D_{12},~C_2,~D_{21}\}.$
\FS{The nonlinear state-space model~\eqref{eq:state-space-model}} represents a wide class of networks; in particular, \glspl{rnn} can be recovered by choosing $A=B=C=D=0$, $B_2=I$, and adding bias terms, as seen in, for example, in ~\citet{goodfellow2016deep}. \FS{Thus, we call model~\eqref{eq:state-space-model} a dynamic \gls{rnn}.}

In this work, we are interested in a regional form of stability.
Hereto, we defined regional stability as follows.
\begin{defn}[Regional stability]
    \label{dfn:regional_stability}
    Consider the compact set $\Xc\subset \Rbb^n$ and a \DF{set of admissible input trajectories $\Uc \subset \ell_{2e}^r$.}
    The model $\Sc_\theta$ in \eqref{eq:state-space-model} is regionally stable on $\Xc$ for $\Uc$ if for any $x_0\in \Xc$ and input sequence from $\Uc$, the set $\Xc$ is forward invariant under the dynamics of $\Sc_\theta$.
\end{defn}

The problem of identifying a model \eqref{eq:state-space-model} that is regionally-stable given a data set \eqref{eq:dataset} is formalized as follows.

\begin{prob}
    \label{prob:regional-sys-id}
    Given a dataset $\Dc$ of the form \eqref{eq:dataset}, find the parameters $\theta$ of the model $\Sc_\theta$ in~\eqref{eq:state-space-model} such that 
    $\Sc_\theta$ minimizes the \gls{mse}\footnote{\FS{We define the \gls{mse} as quadratic error between the predictions of model \eqref{eq:state-space-model} and the outputs from the dataset~\eqref{eq:dataset}, i.e., $\operatorname{MSE}(\hat \ee, \ee) \coloneqq \frac{1}{N}\sum_{k=0}^{N-1}| \eHatK{k} - \ek{k}|^2$.}} 
    between the predictions $\hat \ee$ and the outputs from the dataset $\ee$ in $\Dc$ and is \emph{regionally-stable} according to Definition~\ref{dfn:regional_stability} for some $\Xc$ and $\Uc$.
\end{prob}

Note that the problem statement does not impose a predefined $\Xc$ and $\Uc$. Instead, these follow from the training procedure.
In the next section, we will review the \emph{standard} and \emph{generalized} sector conditions before stating the main theoretical contribution of the paper,
\FS{which is the key to solving Problem~\ref{prob:regional-sys-id}.}

%% file: fig/model.tex
\begin{tikzpicture}[node distance = 1em and 1em, auto, align=center]

    \node[block] (lin) {
        $G$
    };
    \coordinate[] (input_lin_d) at ($(lin.south east)!0.75!(lin.north east)$);
    \coordinate[] (input_lin_u) at ($(lin.south east)!0.25!(lin.north east)$);
    \coordinate[] (output_lin_e) at ($(lin.south west)!0.75!(lin.north west)$);
    \coordinate[] (output_lin_y) at ($(lin.south west)!0.25!(lin.north west)$);

    \node[block, below = of lin] (delta) {
        $\Delta$
    };
    \node[dotted_block, fit = (lin) (delta)] (rnn) {};
    \node at (rnn.west) [left] {$\Sc_\theta$};

    \draw[<-] (input_lin_d) -- ++(4em,0) node[above] {$\dd$};
    \draw[->] (output_lin_e) -- ++(-4em, 0) node[above] {$\hat \ee$};
    \draw[->] (output_lin_y) -- ++(-1.5em,0) |- (delta.west) node[above left]{$\z$};
    \draw[<-] (input_lin_u) -- ++(1.5em,0) |- (delta.east) node[above right]{$\w$}; 
    

\end{tikzpicture}

%% file: content/analysis.tex
The nonlinear model~\eqref{eq:state-space-model} is an interconnection between a linear model and a static, memoryless, and sector-bounded nonlinear function. In the literature, these models are known as \emph{Lur'e}-type systems; see, for example, \citep{Khalil:1173048,shakib2022computationally, scherer2022dissipativity}. To assess their stability properties, sector conditions are used, as shown in \cite {Khalil:1173048}.
First, we recall \emph{standard} and \emph{generalized} sector conditions, and then present convex constraints to guarantee regional stability of the learned model.

\subsection{Sector-bounded nonlinearities}
The scalar nonlinear function $\psi:\Rbb \rightarrow \Rbb$ in model~\eqref{eq:state-space-model} is said to satisfy \emph{standard} sector conditions globally if the following inequality holds
\DF{\begin{equation}
    \label{eq:sector-bounded}
    \mat{c}{\psi(\z) \\ \z}^\T\mat{cc}{-2 & 1 \\ 1 & 0}\mat{c}{\psi(\z) \\ \z} \geq 0
\end{equation}}
for all $\z \in \Rbb$.
The graph of functions that satisfy the \emph{standard} sector condition lies in the sector defined by two lines with slopes $0$ and $1$.
These bounds are conservative since they generally do not tightly cover the nonlinear shape of $\psi$. 
Therefore, global stability results that rely on standard sector conditions are often conservative too.

\begin{figure}
    \centering
    \input{fig/nonlinearities}
    \vspace{-10pt}
    \caption{Nonlinear activation functions.}
    \vspace{3pt}
    \label{fig:nonlinearities}
\end{figure}
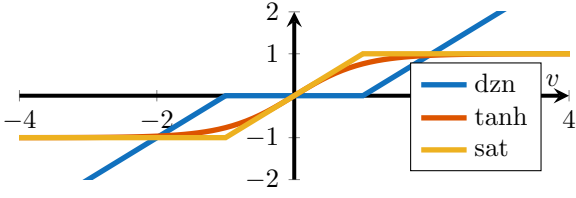

\emph{Lur'e}-type systems, as in~\eqref{eq:state-space-model}, are universal function approximators~\citep{suykens1995artificial}.
Changing the activation function from $\tanh$, which is used in \glspl{rnn}, to saturation ($\sat$) does not affect this property.
Moreover, applying a loop transformation (see~\cite{tarbouriech2006stability}), the \DF{$\operatorname{saturation}$} nonlinearity can be transformed to the deadzone nonlinearity $\psi: \Rbb \mapsto \Rbb$, defined by
\begin{equation}
    \label{eq:dzn2}
     \psi(\z) \coloneqq \dzn(z) \coloneqq
    \begin{cases}
        0 & \mathrm{if}~ |\z|\leq 1, \\ 
        \z-1 & \mathrm{if}~ \z> 1, \\ 
        \z+1 & \mathrm{else}~ \z> 1.
    \end{cases}
\end{equation}
The graphs of the function $\tanh$, $\sat$, and $\dzn$ are displayed in Figure \ref{fig:nonlinearities}.

In model~\eqref{eq:state-space-model}, we use the static decentralized deadzone nonlinear function~\eqref{eq:dzn2} applied elementwise.
Note that the resulting model is still a universal approximator; however, now we can use a state-dependent quadratic function
\begin{equation}
    \label{eq:Gamma}
    \Gamma(\z,\x,\Lambda, H) : = 
    \mat{c}{\Delta(\z) \\ \z + Hx}^\T
    \mat{cc}{-2 \Lambda & \Lambda \\ \Lambda & 0}
    \mat{c}{\Delta(\z) \\ \z + Hx},
\end{equation}
which extends the \emph{standard} sector condition by the following result.
\begin{lem}
    \label{lem:generalized-sector-condition-dzn}
    Consider the nonlinear operator $\Delta:\Rbb^{\nz} \mapsto \Rbb^{\nw}$ 
    with activation function $\psi(z) = \dzn(z)$ applied elementwise,
    and matrices $\Lambda \in \mathbb{D}_+$ and $H \in \mathbb{R}^{\nz \times \nx}$.
    Then, for every $\x \in \Lc(H)$ and $\z \in \mathbb{R}^{\nz}$, we have
    \begin{equation}
    \label{eq:generalized_sector_condition}
        \Gamma(\z,\x,\Lambda,H) \geq 0.
    \end{equation}
\end{lem}

We refer to \eqref{eq:generalized_sector_condition} as the \emph{generalized} sector conditions.
Note that if $H$ is zero, the \emph{standard} sector conditions are recovered. 
However, if $H$ is non-zero, then there is a region around $x$, namely $x\in\Lc(H)$, such that $\Gamma \ge 0$ holds.

\DF{\begin{rem}
    The \emph{generalized} sector conditions are studied in anti-wind-up controller design; see, for example, \citep{tarbouriech2006stability} and references therein. 
\end{rem}}

\subsection{Analysis result}
We will now exploit the properties of \emph{generalized} sector conditions to derive the sets $\Xc$ and $\Uc$ for regional stability. In controller design of neural network models, these properties have been used by \citet{la2025regional} to ensure regional closed-loop stability.
Therein, however, the closed-loop system is an autonomous system, i.e., the set $\Uc$ is the empty set, and $\Xc$ then is a region of attraction of the origin in Definition~\ref{dfn:regional_stability}.
For systems with inputs, the compact set $\Xc$ is no longer a region of attraction; instead, its forward invariance needs to be proven separately and depends on the set of admissible inputs $\Uc$.

In this work, we consider the following static set of inputs:
\begin{equation}
    \label{eq:input-set}
    \Uc_\delta \coloneq \left\{\left(\dk{k} \in \Rbb^{\nd}\right)_{k\in \Nbb} \, \vert \, \dk{k}^\T \dk{k} \le \delta^2, \, \forall\, k \in \Nbb \right\},
\end{equation}
which is defined through the constant $\delta \ge 0$.
The following result provides sufficient conditions for regional stability. It characterizes the positively invariant set, i.e., the region $\Xc$ in the state space and the scalar $\delta$.

\begin{thm}[Regional input-to-state stability]
    \label{thm:regional_iss}
    Given a \newline discrete-time system \eqref{eq:state-space-model} with parameters $\theta$ and decentralized deadzone nonlinearity $\Delta:\Rbb^{\nz} \mapsto \Rbb^{\nw}$, activation function $\psi(\z) = \dzn(\z)$ \eqref{eq:dzn2} applied elementwise,    
    and consider a given scalar $\alpha \in (0,1)$. 
    Suppose there exist matrices $P \in \Sbb_+$, $M\in \Dbb_+$, $L=\mat{ccc}{(l^1)^\T & \cdots & (l)^{\nz})^\T}^\T \in \Rbb^{\nz \times \nx}$, and scalar $s\in\Rbb$ such that the \DF{\glspl{lmi}}
    \begin{subequations}
        \label{eq:thm:conditions}
        \begin{equation}
            \label{eq:thm:lyapunov-lmi}
            \underbrace{\mat{ccc|c}{
                -\alpha^2 P & 0& PC_2^\T  + L^\T & P A^\T \\ 
                \star & -I & D_{21}^\T & B^\T \\ 
                \star & \star & -2M & MB_2^\T \\ \hline  
                \star & \star & \star & -P }
            }_{\Fb} \preceq 0, \, P \succ 0,
        \end{equation}
        \FS{and} 
        \begin{equation}
            \label{eq:thm:regionallity-lmi}
            \underbrace{\mat{cc}{1/s^2 & l^i \\ (l^i)^\T & P}}_{\Gb_i}
            \succeq 0, \qquad \text{for all} \qquad i=1,\ldots,\nz,
        \end{equation}
    \end{subequations}
    are satisfied.
    Then, for any input sequence $\dd \in \mathcal{U}_\delta$ with 
    \begin{equation}\label{eq:thm:delta}
        \delta^2 \le (1-\alpha^2)s^2,
    \end{equation}    
    the ellipsoid $\Ec(P^{-1}/s^2)~\subset~\Lc(H)$ is forward invariant 
    with $H = LP^{-1}$.
    Moreover, for any $\xk{0} \in \Ec(P^{-1}/s^2)$ and input sequence $u\in\mathcal{U}_\delta$, the following bound is satisfied for any $k \ge 0$:
    \begin{multline}\label{eq:thm:ISS}
            |\xk{k}| \le 
            \text{min} 
            \Bigg\{
            \sqrt{\frac{\lambda_\textrm{max}}{\lambda_\textrm{min}}}
            \alpha^{k}|\xk{0}|+\sqrt{\frac{1}{(1-\alpha^2)\lambda_\textrm{min}}}
       \sup_{0 \le j < k} |\dd_k|, \\
       \frac{s}{\sqrt{\lambda_\textrm{min}}}
        \Bigg\}.
        \end{multline}
    where $\lambda_\textrm{min}$ and $\lambda_\textrm{max}$ denote the minimum and maximum eigenvalues of the matrix $P^{-1}$.
\end{thm}

Theorem~\ref{thm:regional_iss} defines the size of input set $\mathcal{U}_\delta$ and the region $\Xc = \Ec(P^{-1}/s^2)$, required for regional stability in Definition~\ref{dfn:regional_stability}.
In addition,~\eqref{eq:thm:ISS} characterizes the size of the state $\xk{k}$, which depends on the size of the input.
This is similar to a regional input-to-state stability result.
In particular, for $\dd=0$, the origin $\x=0$ is a locally exponentially stable equilibrium with the region of attraction~$\Xc$.
In the next section, the \gls{lmi} constraints in Theorem~\ref{thm:regional_iss} will be imposed during model training to ensure that the trained model is regionally stable.

\begin{rem}
    \label{rem:iss}
    Global stability is recovered for the case $H=0$ ($L=0$). In this case, we see from \eqref{eq:thm:delta} that $\delta$ can be arbitrarily large since \eqref{eq:thm:regionallity-lmi} holds for any $s \geq0$. The sets $\Xc$ is then the entire space $\Rbb^{\nx \times \nx}$. Furthermore, then \eqref{eq:sector-bounded} implies a global input-to-state stability bound.
\end{rem}


%% file: fig/nonlinearities.tex
%
%
\definecolor{mycolor1}{rgb}{0.06600,0.44300,0.74500}%
\definecolor{mycolor2}{rgb}{0.86600,0.32900,0.00000}%
\definecolor{mycolor3}{rgb}{0.92900,0.69400,0.12500}%
\definecolor{mycolor4}{rgb}{0.12941,0.12941,0.12941}%
\begin{tikzpicture}

\begin{axis}[%
width=\columnwidth,
height=3.8cm,
at={(2.6in,1.013in)},
xmin=-4,
xmax=4,
xlabel style={font=\color{mycolor4}},
xlabel={$\z$},
ymin=-2,
ymax=2,
ylabel style={font=\color{mycolor4}},
axis background/.style={fill=white},
title style={font=\bfseries\color{mycolor4}},
axis x line=middle,
axis y line=middle,
axis line style={-stealth, line width=1.5pt},
legend style={legend cell align=left, align=left, at={(0.95,0.05)}, anchor=south east}
]
\addplot [color=mycolor1, line width=2.0pt]
  table[row sep=crcr]{%
-5	-4\\
-4.9	-3.9\\
-4.8	-3.8\\
-4.7	-3.7\\
-4.6	-3.6\\
-4.5	-3.5\\
-4.4	-3.4\\
-4.3	-3.3\\
-4.2	-3.2\\
-4.1	-3.1\\
-4	-3\\
-3.9	-2.9\\
-3.8	-2.8\\
-3.7	-2.7\\
-3.6	-2.6\\
-3.5	-2.5\\
-3.4	-2.4\\
-3.3	-2.3\\
-3.2	-2.2\\
-3.1	-2.1\\
-3	-2\\
-2.9	-1.9\\
-2.8	-1.8\\
-2.7	-1.7\\
-2.6	-1.6\\
-2.5	-1.5\\
-2.4	-1.4\\
-2.3	-1.3\\
-2.2	-1.2\\
-2.1	-1.1\\
-2	-1\\
-1.9	-0.9\\
-1.8	-0.8\\
-1.7	-0.7\\
-1.6	-0.6\\
-1.5	-0.5\\
-1.4	-0.4\\
-1.3	-0.3\\
-1.2	-0.2\\
-1.1	-0.0999999999999996\\
-1	0\\
-0.899999999999999	0\\
-0.8	0\\
-0.7	0\\
-0.6	0\\
-0.5	0\\
-0.399999999999999	0\\
-0.3	0\\
-0.199999999999999	0\\
-0.0999999999999996	0\\
0	0\\
0.0999999999999996	0\\
0.199999999999999	0\\
0.3	0\\
0.399999999999999	0\\
0.5	0\\
0.6	0\\
0.7	0\\
0.8	0\\
0.899999999999999	0\\
1	0\\
1.1	0.0999999999999996\\
1.2	0.2\\
1.3	0.3\\
1.4	0.4\\
1.5	0.5\\
1.6	0.6\\
1.7	0.7\\
1.8	0.8\\
1.9	0.9\\
2	1\\
2.1	1.1\\
2.2	1.2\\
2.3	1.3\\
2.4	1.4\\
2.5	1.5\\
2.6	1.6\\
2.7	1.7\\
2.8	1.8\\
2.9	1.9\\
3	2\\
3.1	2.1\\
3.2	2.2\\
3.3	2.3\\
3.4	2.4\\
3.5	2.5\\
3.6	2.6\\
3.7	2.7\\
3.8	2.8\\
3.9	2.9\\
4	3\\
4.1	3.1\\
4.2	3.2\\
4.3	3.3\\
4.4	3.4\\
4.5	3.5\\
4.6	3.6\\
4.7	3.7\\
4.8	3.8\\
4.9	3.9\\
5	4\\
};
\addlegendentry{dzn}

\addplot [color=mycolor2, line width=2.0pt]
  table[row sep=crcr]{%
-5	-0.999909204262595\\
-4.9	-0.999889102950554\\
-4.8	-0.99986455170076\\
-4.7	-0.999834565554297\\
-4.6	-0.999797941612184\\
-4.5	-0.999753210848027\\
-4.4	-0.999698579283881\\
-4.3	-0.999631856190073\\
-4.2	-0.999550366459533\\
-4.1	-0.999450843687797\\
-4	-0.999329299739067\\
-3.9	-0.999180865670028\\
-3.8	-0.998999597785841\\
-3.7	-0.998778241281131\\
-3.6	-0.998507942332327\\
-3.5	-0.998177897611199\\
-3.4	-0.997774927934279\\
-3.3	-0.997282960099142\\
-3.2	-0.996682397839651\\
-3.1	-0.9959493592219\\
-3	-0.99505475368673\\
-2.9	-0.993963167350583\\
-2.8	-0.992631520201128\\
-2.7	-0.991007453678118\\
-2.6	-0.989027402201099\\
-2.5	-0.98661429815143\\
-2.4	-0.98367485769368\\
-2.3	-0.980096396266191\\
-2.2	-0.975743130031452\\
-2.1	-0.970451936613454\\
-2	-0.964027580075817\\
-1.9	-0.956237458127739\\
-1.8	-0.946806012846268\\
-1.7	-0.935409070603099\\
-1.6	-0.921668554406471\\
-1.5	-0.905148253644866\\
-1.4	-0.885351648202263\\
-1.3	-0.861723159313306\\
-1.2	-0.833654607012155\\
-1.1	-0.80049902176063\\
-1	-0.761594155955765\\
-0.899999999999999	-0.716297870199024\\
-0.8	-0.664036770267849\\
-0.7	-0.604367777117164\\
-0.6	-0.537049566998035\\
-0.5	-0.46211715726001\\
-0.399999999999999	-0.379948962255224\\
-0.3	-0.291312612451591\\
-0.199999999999999	-0.197375320224903\\
-0.0999999999999996	-0.0996679946249555\\
0	0\\
0.0999999999999996	0.0996679946249555\\
0.199999999999999	0.197375320224903\\
0.3	0.291312612451591\\
0.399999999999999	0.379948962255224\\
0.5	0.46211715726001\\
0.6	0.537049566998035\\
0.7	0.604367777117164\\
0.8	0.664036770267849\\
0.899999999999999	0.716297870199024\\
1	0.761594155955765\\
1.1	0.80049902176063\\
1.2	0.833654607012155\\
1.3	0.861723159313306\\
1.4	0.885351648202263\\
1.5	0.905148253644866\\
1.6	0.921668554406471\\
1.7	0.935409070603099\\
1.8	0.946806012846268\\
1.9	0.956237458127739\\
2	0.964027580075817\\
2.1	0.970451936613454\\
2.2	0.975743130031452\\
2.3	0.980096396266191\\
2.4	0.98367485769368\\
2.5	0.98661429815143\\
2.6	0.989027402201099\\
2.7	0.991007453678118\\
2.8	0.992631520201128\\
2.9	0.993963167350583\\
3	0.99505475368673\\
3.1	0.9959493592219\\
3.2	0.996682397839651\\
3.3	0.997282960099142\\
3.4	0.997774927934279\\
3.5	0.998177897611199\\
3.6	0.998507942332327\\
3.7	0.998778241281131\\
3.8	0.998999597785841\\
3.9	0.999180865670028\\
4	0.999329299739067\\
4.1	0.999450843687797\\
4.2	0.999550366459533\\
4.3	0.999631856190073\\
4.4	0.999698579283881\\
4.5	0.999753210848027\\
4.6	0.999797941612184\\
4.7	0.999834565554297\\
4.8	0.99986455170076\\
4.9	0.999889102950554\\
5	0.999909204262595\\
};
\addlegendentry{tanh}

\addplot [color=mycolor3, line width=2.0pt]
  table[row sep=crcr]{%
-5	-1\\
-4.9	-1\\
-4.8	-1\\
-4.7	-1\\
-4.6	-1\\
-4.5	-1\\
-4.4	-1\\
-4.3	-1\\
-4.2	-1\\
-4.1	-1\\
-4	-1\\
-3.9	-1\\
-3.8	-1\\
-3.7	-1\\
-3.6	-1\\
-3.5	-1\\
-3.4	-1\\
-3.3	-1\\
-3.2	-1\\
-3.1	-1\\
-3	-1\\
-2.9	-1\\
-2.8	-1\\
-2.7	-1\\
-2.6	-1\\
-2.5	-1\\
-2.4	-1\\
-2.3	-1\\
-2.2	-1\\
-2.1	-1\\
-2	-1\\
-1.9	-1\\
-1.8	-1\\
-1.7	-1\\
-1.6	-1\\
-1.5	-1\\
-1.4	-1\\
-1.3	-1\\
-1.2	-1\\
-1.1	-1\\
-1	-1\\
-0.899999999999999	-0.899999999999999\\
-0.8	-0.8\\
-0.7	-0.7\\
-0.6	-0.6\\
-0.5	-0.5\\
-0.399999999999999	-0.399999999999999\\
-0.3	-0.3\\
-0.199999999999999	-0.199999999999999\\
-0.0999999999999996	-0.0999999999999996\\
0	0\\
0.0999999999999996	0.0999999999999996\\
0.199999999999999	0.199999999999999\\
0.3	0.3\\
0.399999999999999	0.399999999999999\\
0.5	0.5\\
0.6	0.6\\
0.7	0.7\\
0.8	0.8\\
0.899999999999999	0.899999999999999\\
1	1\\
1.1	1\\
1.2	1\\
1.3	1\\
1.4	1\\
1.5	1\\
1.6	1\\
1.7	1\\
1.8	1\\
1.9	1\\
2	1\\
2.1	1\\
2.2	1\\
2.3	1\\
2.4	1\\
2.5	1\\
2.6	1\\
2.7	1\\
2.8	1\\
2.9	1\\
3	1\\
3.1	1\\
3.2	1\\
3.3	1\\
3.4	1\\
3.5	1\\
3.6	1\\
3.7	1\\
3.8	1\\
3.9	1\\
4	1\\
4.1	1\\
4.2	1\\
4.3	1\\
4.4	1\\
4.5	1\\
4.6	1\\
4.7	1\\
4.8	1\\
4.9	1\\
5	1\\
};
\addlegendentry{sat}

\end{axis}
\end{tikzpicture}%

%% file: content/training.tex
During training, we use the input-output dataset~\eqref{eq:dataset} to optimize the parameters $\theta$ of the system \eqref{eq:state-space-model}. In addition, the matrices $P$, $M$, and $L$, as well as the scalars $s$ and $\alpha$ are optimized. Thus, the set of learnable parameters consists of $\omega = \{\theta,~P,~L,~M,~\alpha,~s\}.$
The objective of the optimization problem is to reduce the error between the prediction made by the model and the outputs in the training dataset
\begin{equation}
    \label{eq:train:opt}
    \begin{aligned}
        \min_\omega  && &| \Sc_\theta(\dd, \xk{0}) - \ee |^2 \\
        \mathrm{such \ that} && & \eqref{eq:thm:conditions}~\mathrm{and}~ \eqref{eq:thm:delta}~ \mathrm{hold, and \ } 0<\alpha<1.
    \end{aligned}
\end{equation}
\DF{By the function $\Sc_\theta(\dd, \xk{0})$, we denote the predicted output response of model \eqref{eq:state-space-model} for input $\dd$ and initial condition $\xk{0}$.} 
Due to the recurrence of the model \eqref{eq:state-space-model}, the optimization problem \eqref{eq:train:opt} is nonlinear. The constraints, however, are affine with respect to the parameters $P, L, M, s^2$, which will be useful during initialization and training.

\subsection{Interior-point method}
We solve the optimization problem \eqref{eq:train:opt} iteratively using the Adam optimizer \citep{kingma2014adam}, a stochastic gradient-descent method with momentum.
To satisfy the conditions of \eqref{eq:train:opt}, we utilize barrier functions, known from interior-point methods (see~\citet{boyd2004convex}).
Intuitively, parameters close to the constraint boundary receive a large loss such that the resulting gradient does not point towards infeasible parameters. Given a constraint $\Cb\prec 0$, the associated barrier function is defined as
\begin{equation}
    \label{eq:barrier_function}
    \phi(\Cb) \vcentcolon = \begin{cases}
        -\log \operatorname{det}(-\Cb) & \text{ if } \Cb \prec 0,\\
        \infty & \text{ if } \Cb \nprec 0.
    \end{cases}
\end{equation}
and it is added to the \gls{mse} between the prediction and the outputs from the training dataset. Thus, the training loss is defined as
\DF{\begin{align}
        \Lb(\ee, (\dd, \xk{0}))  =& \frac{1}{N}|\Sc_\theta(\dd, \xk{0}) - \ee|^2+ \notag \\ 
        & \nu \Bigl(\phi(\Fb) + \sum_{i=1}^{\nz} \phi(-\Gb_i) \Bigr. \label{eq:training-loss}\\ 
        & \Bigl. + \phi(\delta^2 - (1-\alpha^2)s^2)- \phi(1-\alpha) - \phi(\alpha)\Bigr) \notag.
\end{align}}%
The matrices $\Fb$ and $\Gb_i$ stem from conditions \eqref{eq:thm:conditions} and the last line of~\eqref{eq:training-loss} refers to the scalar inequality \eqref{eq:thm:delta} and ensures that $\alpha \in (0,~1)$. The regularization parameter $\nu$ controls the influence of the barrier functions and can be scheduled during training.

The regularization is a soft constraint that does not guarantee that a new set of parameters satisfies the constraints \eqref{eq:thm:conditions} and \eqref{eq:thm:delta}. Therefore, after every epoch, we check these conditions using a Cholesky decomposition of the matrices $-\Fb$ and $\Gb_i$, and we verify scalar inequality \eqref{eq:thm:delta} and the condition $\alpha \in (0,1)$.
In the case that the constraints are not satisfied after a parameter update, we fix the model parameters $\theta$, $s$, $\alpha$, and try to find $P$, $L$, $M$ to satisfy conditions \eqref{eq:thm:conditions} and \eqref{eq:thm:delta}. This is a feasibility problem that can be formulated as an \acrlong{sdp} and solved efficiently with off-the-shelf solvers. If no feasible solution is found, the parameters are rolled back to the values before the update.


\subsection{Initialization}
To initialize the optimization problem, we aim to find a feasible initial model such that the barrier functions in the loss function~\eqref{eq:training-loss} are bounded.
The next proposition presents a method for finding such a feasible initial model.

\begin{prop}
    \label{prop:init_params}
    Take $\alpha=0.99,~A=0.9I,~C_2 = \Rc(-1,1), ~C = [I, 0],$ and $B_2 =D = D_{12} = 0$.
    Furthermore, given a user-specified $\delta >0$, take $s = \sqrt{\frac{\delta^2}{(1-\alpha^2)}}$.    
    If there exists parameters $P,L,M,D_{21}$, and $B$ such that inequality~\eqref{eq:thm:conditions} is satisfied, then model~\eqref{eq:state-space-model} satisfies all condition of Theorem~\ref{thm:regional_iss}.
    Furthermore, given a $\beta > 0$, if the \gls{lmi}
    \begin{equation}
        \label{eq:prop:init_cond}
         \beta^2 I \preceq s^2 P    
    \end{equation}
    is satisfied, then any $x_0\in\Rbb^{\nx}$ that satisfies
    $|\xk{0}|^2 \le \beta^2$ is included in the set $\Ec(P^{-1}/s^2)$. 
\end{prop}

Proposition \ref{prop:init_params} ensures a feasible set of initial parameters, and thus the employed barrier functions are bounded.
Furthermore, if additional information is available about the initial conditions that should be in the region $\Xc$, then the additional \gls{lmi} constraint~\eqref{eq:prop:init_cond} can be used to ensure that a ball of radius $\beta$ is contained in the region $\Xc$.

\subsection{Post processing}
After the training algorithm is finished, we analyze the learned model. \DF{We fix $\theta$ and $\alpha$ and optimize for a large compact set defined by $\Xc$ under the constraints given by the generalized sector conditions $\Ec(P^{-1}/s^2) \subset \Lc(H)$. This is achieved by minimizing $\hat s = 1/\sqrt{s}$ which appears in the following \acrlong{sdp}}
\begin{equation}
    \label{eq:post:opt-s}
    \begin{aligned}
    \min_{\hat s, P, L, M} && &\hat{s} \\
    \mathrm{such\ that} && & \eqref{eq:thm:lyapunov-lmi},~\smat{\hat s & l^i \\ (l^i)^\T & P}\preceq 0,~i=1,\ldots m, ~\mathrm{and}~\eqref{eq:thm:delta}.
    \end{aligned}
\end{equation}
Then we get $s = \frac{1}{\sqrt{\hat s}}$, and $\Ec(P^{-1}/s^2) \subset \Lc(H)$. 

%% file: content/evaluation.tex
First, we introduce the unknown data-generating system and the baseline models. Then, we discuss the results and compare our method against two baselines.


\subsection{Unknown data-generating system}
To demonstrate the advantage of using generalized sector conditions in system identification with \glspl{rnn}, we consider a synthetic nonlinear system with two states, i.e., $n=2$, and a deadzone nonlinearity with $m=2$.
The state-space description in \eqref{eq:state-space-model} is characterized by the matrices
\begin{equation}
\label{eq:example:system}
    \smat{
        A & B & B_2 \\ 
        C & D & D_{12} \\ 
        C_2 & D_{21} & 0
    } = 
    \smat{
        0.998 & 0.096 & 0.0049 & 0.4191 & 0.4191 \\
        -0.048 & 0.921 & 0.096 & 0.3744 &   0.3744 \\ 
        1 & 0 & 0 & 1 & 1 \\ 
        0.18 &   0 & 1 & 0 & 0 \\
        0 & 0.18 & 1 & 0 & 0
    }   
\end{equation}
and deadzone nonlinear function $\psi = \dzn$.
The system \eqref{eq:example:system} satisfies all the conditions of Theorem \ref{thm:regional_iss} for a non-zero $L$ and is regionally-stable according to Definition~\ref{dfn:regional_stability} (see Remark \ref{rem:iss}). 
The set of inputs is as in \eqref{eq:input-set} with \DF{$\delta = \max_{k,i}|(\dk{k})^{[i]}|$} 
We used the system \eqref{eq:example:system} to generate a dataset of the form \eqref{eq:dataset}. As inputs, we use
\DF{
\begin{enumerate}[label=(\roman*)]
    \item $\dd = \sqrt{s^2 (1-\alpha^2)} \sin(k dt)$ \label{enum:sin}
    \item $\dd = \Rc(-\sqrt{s^2 (1-\alpha^2)}, ~\sqrt{s^2 (1-\alpha^2)})$ \label{enum:noise},
\end{enumerate}
}
where $s$ defines the size of the ellipsoidal region $\Xc$, $\alpha=0.97$ is the contraction rate and $dt=0.1$ refers to the sampling time. 
We used random initial conditions in the range $\xk{0}^1, \xk{0}^2 \in [-6,~6]$ and iterated the system for $N_i=N=50$, steps for all $i$. \FS{We generated $M_{\sin} = 300$ samples with sinusoidal input (see \ref{enum:sin}) and $M_{\mathrm{noise}}=300$ trajectories with noise input (see \ref{enum:noise}). In addition, we added $M_{\sin,~\mathrm{zero}}=M_{\mathrm{noise,~zero}}=150$ trajectories that start from $\xk{0} = [0,~0]^\T$.} Thus, the input-output dataset consists of $45~000$ datapoints and $900$ trajectories. 
Note that some initial conditions are not in the set $\Xc$. Hence, some trajectories are in fact diverging.


We compare our model (\MGenSec{}) against a model that uses standard sector conditions (\MStdSec{}) and achieves global stability results in the form of input-to-state stability (see Remark \ref{rem:iss}), and against a model without constraints during learning (\MLtiRnn{}). 
The experiments in this work were conducted on an Intel i7 Laptop, and each model took approximately $50$ minutes to train for $4000$ epochs. 
The code can be found on GitHub\footnote{\url{https://github.com/Dany-L/genSecSysId}}.

\begin{figure*}[t]
        \centering
        \begin{subfigure}{0.33\textwidth}
            \input{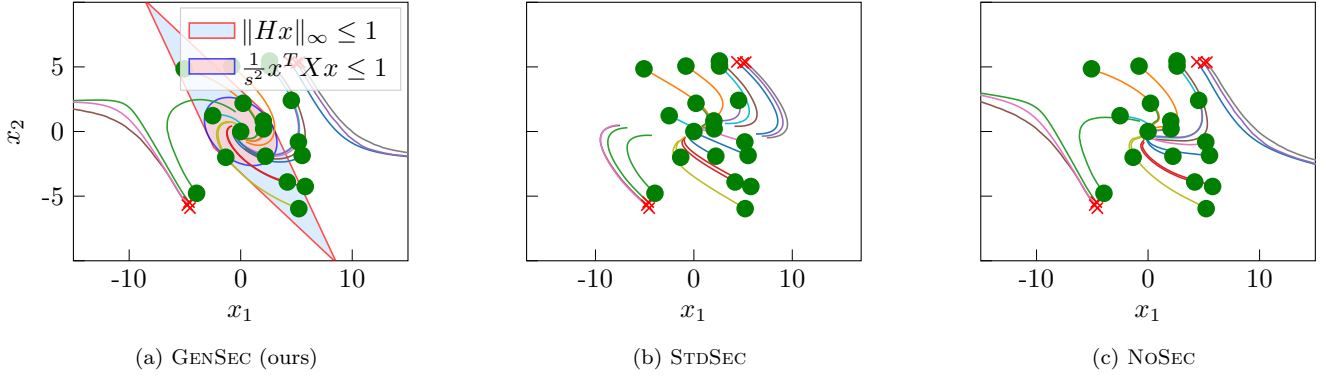}
            \caption{\MGenSec{}}
            \label{fig:GenSec-phase-space}
        \end{subfigure}
        \begin{subfigure}{0.33\textwidth}
            \centering
            \input{fig/stdSec-phase-space}
            \caption{\MStdSec{}}
            \label{fig:stdSec-phase-space}
        \end{subfigure}%
        \begin{subfigure}{0.33\textwidth}
            \centering
            \input{fig/ltiRnn-phase-space}
            \caption{\MLtiRnn{}}
            \label{fig:ltiRnn-phase-space}
        \end{subfigure}
        \vspace{-15pt}
        \caption{Predictions made by the learned models in phase space, with excitation from the test dataset. The red crosses indicate initial conditions of trajectories that are diverging in the dataset, while the green dots indicate the ones that end up in $\Xc$. \FS{In \subref{fig:GenSec-phase-space}, we see that the predictions made by our model are consistent with the dataset in the sense that when the trajectories from the dataset diverge, also the predicted trajectories diverge.} In addition, we see the learned compact sets $\Ec(X/s^2)$ and $\Lc(H)$ in \subref{fig:GenSec-phase-space}. The predictions of \MStdSec{} in \subref{fig:stdSec-phase-space} show that, even if the true system trajectories diverge, the models still predict converging responses, which is inconsistent and results from the imposed global stability conditions. The predictions of model \MLtiRnn{} in \subref{fig:ltiRnn-phase-space} are consistent but do not provide sets $\Xc$ or $\Uc_\delta$. }
        
    \end{figure*}   


\begin{figure*}
    \centering
    \input{fig/comparison/outputs.tex}
    \vspace{-10pt}
    \caption{\FS{Predictions of the models in the time domain compared to the outputs from the test dataset.}}
    \label{fig:eval:outputs}
\end{figure*}
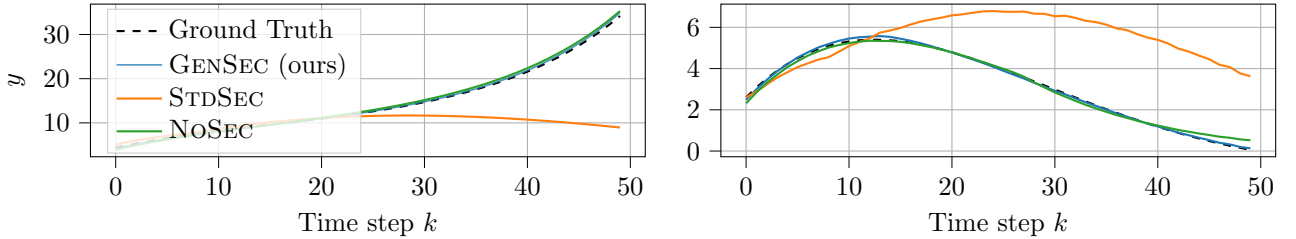

\subsection{Discussion of results}
In Figure \ref{fig:GenSec-phase-space}, we see that the learned model \MGenSec{} is regionally stable, which is consistent with the data-generating system. Trajectories that diverge in the data-generating system also diverge for the learned model, and trajectories that end up in $\Xc$ also end up in the compact set. The model \MStdSec{} is globally input-to-state stable, as evidenced by the converging trajectories in Figure \ref{fig:stdSec-phase-space}, even when the true system trajectories diverge, which is inconsistent. When we do not add any constraints (the model \MLtiRnn{}) on the learnable parameters, we have no stability guarantees and, consistent with the data-generating system, observe diverging trajectories, as shown in Figure \ref{fig:ltiRnn-phase-space}. This underpins the theoretical results of Theorem \ref{thm:regional_iss}, showing that with generalized sector conditions, we can recover the regional stability properties of the true system. With the proposed method, compared to global approaches, we extend the class of models that we can recover and achieve \FS{behavior that is consistent with the data-generating system.}

In Figure \ref{fig:eval:outputs}, this is shown in the time domain. The plot on the left shows a diverging trajectory. While \MGenSec{} and \MLtiRnn{} can recover such diverging behavior, the constraints in \MStdSec{} do not allow for regionally stable models. The model \MGenSec{} can recover converging and diverging trajectories from the training dataset, as shown in Figure \ref{fig:eval:outputs}. It also provides the compact set $\Xc$ and the input set $\Uc_\delta$ for reliable model use. The final loss values on the test dataset are shown in Table \ref{tab:evaluation}, \DF{and confirm that \MGenSec{} outperforms the baseline models in terms of prediction accuracy}.
\begin{table}
    \centering
    \caption{Evaluation results on test dataset for the different models.}
    \label{tab:evaluation}
    \begin{tabular}{llll}
        \toprule
        Model & NRMSE & \# params & $\nz$ \\
        \midrule
        \MLtiRnn{} & 0.001581 & 24  & 2\\ 
        \MStdSec{} & 0.01635 & 33  & 2\\ 
        \MGenSec{} & \textbf{0.0005698} & 37 & 2 \\ 
        \bottomrule
    \end{tabular}
\end{table}

%% file: fig/stdSec-phase-space.tex
\begin{tikzpicture}

\definecolor{crimson2143940}{RGB}{214,39,40}
\definecolor{darkgray176}{RGB}{176,176,176}
\definecolor{darkorange25512714}{RGB}{255,127,14}
\definecolor{darkturquoise23190207}{RGB}{23,190,207}
\definecolor{forestgreen4416044}{RGB}{44,160,44}
\definecolor{goldenrod18818934}{RGB}{188,189,34}
\definecolor{gray127}{RGB}{127,127,127}
\definecolor{green01270}{RGB}{0,127,0}
\definecolor{mediumpurple148103189}{RGB}{148,103,189}
\definecolor{orchid227119194}{RGB}{227,119,194}
\definecolor{sienna1408675}{RGB}{140,86,75}
\definecolor{steelblue31119180}{RGB}{31,119,180}

\begin{axis}[
    width=\textwidth,
    height=5cm,
    yticklabels={,,},
    xticklabels={,,},
tick align=inside,
tick pos=left,
x grid style={darkgray176},
xlabel={$x_1$},
xmin=-17, xmax=17,
xtick style={color=black},
y grid style={darkgray176},
xticklabels={,-10,0,10},
ymin=-10, ymax=10,
ytick style={color=black}
]
\addplot [semithick, red, mark=x, mark size=3, mark options={solid}, only marks]
table {%
4.37444990644978 5.38906124561606
};
\addplot [semithick, steelblue31119180]
table {%
4.37444990644978 5.38906124561606
4.84333453495579 5.05578772002071
5.26615856792845 4.74421096997507
5.64575648462573 4.45308870493848
5.9847927364234 4.18125094233756
6.28577119287159 3.92759598920172
6.55104417124961 3.69108661025428
6.78282107080275 3.47074637301523
6.98317663067969 3.26565616132072
7.16788330705209 3.0690561432341
7.33946757693819 2.87983537609505
7.49824020608421 2.69782776301867
7.64449996824853 2.52287232356756
7.77853442595185 2.35481293309067
7.90062074079109 2.19349805277101
8.01102650687136 2.03878045231109
8.1100106006482 1.89051692727331
8.19782404027718 1.7485680131597
8.27471084744258 1.61279769836228
8.34090890458258 1.48307313814021
8.39665080044485 1.35926437178285
8.44216465699451 1.24124404509961
8.47767493085462 1.1288871403367
8.50340318268614 1.02207071555996
8.51956880820745 0.920673655460242
8.52638972491036 0.824576435436383
8.52408300894601 0.733660900690002
8.51286547712617 0.647810061928064
8.49295420950901 0.566907909114942
8.46456700860709 0.490839244546651
8.42792279186444 0.419489536338028
8.38324191469265 0.352744793220636
8.33074642202626 0.290491461346976
8.27066022704926 0.232616343587264
8.20320921644935 0.179006541590634
8.12862128226923 0.129549420665358
8.04712628113533 0.0841325973146127
7.95895592234931 0.0426439490477845
7.8643435870173 0.00497164587430216
7.76352408106063 -0.0289957973201943
7.65673332559204 -0.0593694485149562
7.54420798874809 -0.0862598676016768
7.42618506363344 -0.109776998105532
7.30290139755261 -0.130030043111125
7.17459317817277 -0.147127327073983
7.04149538267325 -0.161176145330841
6.9038411962892 -0.172282603236606
6.76186140694534 -0.180551446951398
6.61578378289713 -0.186085887976554
6.4658324404497 -0.188987423593085
6.31222720890735 -0.189355655389388
};
\addplot [semithick, green01270, mark=*, mark size=3, mark options={solid}, only marks]
table {%
-5.07318404917714 4.85075096054016
};
\addplot [semithick, darkorange25512714]
table {%
-5.07318404917714 4.85075096054016
-4.25590069340863 4.49165396508724
-3.50199044218035 4.15831082206014
-2.7735737644947 3.83757359369315
-2.13990342701877 3.5522678273534
-1.59627929970851 3.3002630200097
-1.12926668490057 3.07681348067637
-0.730181961395413 2.87881318330192
-0.403274189660387 2.70758455730573
-0.122964441066759 2.55426409302002
0.114108697439014 2.4175289231463
0.327691158337819 2.28979314132067
0.516496953865255 2.17136756663612
0.679002629470759 2.06263488965753
0.822916128928872 1.96120840595968
0.947898290434967 1.86712572563384
1.05232112852942 1.78081247105405
1.14569903967269 1.69935109591739
1.22261848241712 1.62431444520101
1.29131819938687 1.55317492495949
1.34255779991449 1.48865927872575
1.39292040509358 1.42573789674122
1.42865656198054 1.36850058716933
1.45919580926826 1.31406933399429
1.48027804575609 1.26368452427485
1.5016313177252 1.21438617239464
1.5183303982987 1.16762193183508
1.52493815482254 1.12498874071571
1.52652510125321 1.08492613724037
1.52504364462615 1.04681546164982
1.52231503072883 1.01008027934991
1.5191763695815 0.97444209416912
1.50559478729745 0.942885377400477
1.49580481569678 0.911103797857689
1.47930577305891 0.882228119286344
1.45861267740482 0.855473587525158
1.4384927747286 0.829383080666453
1.4144541340798 0.805283189340867
1.39868566531341 0.779491552798546
1.37434247754371 0.757050259186837
1.3469589083757 0.736269720349736
1.32673547948112 0.714066797048957
1.30059096507097 0.694354850357517
1.27977881422985 0.673732734997111
1.25672572782755 0.654460285909107
1.23748063318721 0.634702672145609
1.21606687870319 0.616239943011999
1.18798091283542 0.600404331651794
1.16857621135071 0.582566079940055
1.14622798035486 0.566206359531807
1.12581187892586 0.5498437681119
};
\addplot [semithick, red, mark=x, mark size=3, mark options={solid}, only marks]
table {%
-4.63963370552342 -5.45600729794191
};
\addplot [semithick, forestgreen4416044]
table {%
-4.63963370552342 -5.45600729794191
-5.12305193514259 -5.11393338572242
-5.55718618653894 -4.79479587574683
-5.9450451879438 -4.49728167342704
-6.2894785876495 -4.22014700130465
-6.59318567343171 -3.96221363230385
-6.8587235109723 -3.72236536356894
-7.08851453480022 -3.49954471691876
-7.28731648537868 -3.29169969763557
-7.47131695398363 -3.09190599315582
-7.64095485200303 -2.89995643685514
-7.79666511402518 -2.71564644967842
-7.93887828762351 -2.53877413382987
-8.06802008106825 -2.36914038051228
-8.18451087520165 -2.20654898983991
-8.28876520603833 -2.05080680094313
-8.38119122491325 -1.90172383019559
-8.46219014319202 -1.75911341542842
-8.53215566868209 -1.62279236394973
-8.59147344093549 -1.49258110216443
-8.64052047261533 -1.36830382458635
-8.67966460400825 -1.24978864005465
-8.709263977605 -1.13686771300742
-8.72966653944274 -1.02937739772791
-8.7412095736073 -0.927158363561432
-8.74421927593559 -0.830055709204118
-8.73901037253959 -0.737919064285802
-8.72588578829971 -0.650602676608475
-8.70513636995019 -0.56796548355683
-8.6770406678086 -0.489871166367169
-8.64186477959084 -0.416188186123594
-8.59986225910844 -0.346789800544136
-8.55127409197296 -0.281554060822402
-8.49632873973921 -0.220363788000334
-8.43524225321251 -0.163106528562757
-8.36821845493152 -0.10967448916239
-8.29544919012488 -0.0599644506026581
-8.2171146447342 -0.0138776614228933
-8.13338372840454 0.028680288355878
-8.04441451967331 0.0677996125585219
-7.95035476994695 0.103566492290703
-7.85134246224707 0.136063256324859
-7.74750642014063 0.165368577102541
-7.63896696174836 0.191557680853445
-7.52583659325585 0.214702570181652
-7.40822073593865 0.234872257336448
-7.28621848035947 0.252133006269854
-7.15992336110628 0.266548581486988
-7.02942414521697 0.278180501619304
-6.89480562728271 0.287088295595651
-6.75614942413782 0.29332975925228
};
\addplot [semithick, green01270, mark=*, mark size=3, mark options={solid}, only marks]
table {%
5.78886904899375 -4.25151721984619
};
\addplot [semithick, crimson2143940]
table {%
5.78886904899375 -4.25151721984619
5.08725868162252 -3.95408172352278
4.42465680524268 -3.67321120770297
3.79893595044292 -3.40799794683774
3.20807487484849 -3.15758051029799
2.65068136547527 -2.92131367632612
2.17321077524922 -2.71416844824998
1.76503678543832 -2.53238820528341
1.41691390205434 -2.37270160397258
1.11664739188002 -2.2304897887245
0.849068554475729 -2.09975136991079
0.610973802569631 -1.97944805242448
0.39945752146052 -1.86863625905144
0.2118847043019 -1.76645847395402
0.0458661746230382 -1.67213540229544
-0.100763850899121 -1.58495886818997
-0.229968083605618 -1.50428538035556
-0.343524571420661 -1.42953030138349
-0.443043379427544 -1.36016256248462
-0.52998170991511 -1.29569987098023
-0.605657613742948 -1.23570436272712
-0.671262430657804 -1.17977865614744
-0.727872083244 -1.12756226861402
-0.776457337389729 -1.07872835965954
-0.817893131393445 -1.03298076886536
-0.852967066024918 -0.990051319373598
-0.882387138908134 -0.949697360781544
-0.906788798430978 -0.911699527744272
-0.926741384939936 -0.875859692952318
-0.942754020184448 -0.841999095285699
-0.955280999778676 -0.809956625891346
-0.964726737797721 -0.779587256704477
-0.971450307475099 -0.750760597549655
-0.975769617277477 -0.723359569427679
-0.977965257364275 -0.697279182931534
-0.978284047560342 -0.672425411949138
-0.976942314449244 -0.648714153912216
-0.974128922005527 -0.62607026884789
-0.970008077301663 -0.604426690390573
-0.964721930226804 -0.583723602723562
-0.958392983819209 -0.563907678149049
-0.951126329723393 -0.544931370638045
-0.943011721419272 -0.526752261293582
-0.934125496217746 -0.509332452176606
-0.924532355560197 -0.492638005399032
-0.914287011884465 -0.476638424786873
-0.903435709213677 -0.461306177762493
-0.892017623674547 -0.446616255392636
-0.880066149346609 -0.43254576880192
-0.867610074172013 -0.419073580363217
-0.854674650106304 -0.406179968250523
};
\addplot [semithick, red, mark=x, mark size=3, mark options={solid}, only marks]
table {%
4.99558976766198 5.30797448402492
};
\addplot [semithick, mediumpurple148103189]
table {%
4.99558976766198 5.30797448402492
5.46559826624032 4.97205310036385
5.88102128868342 4.66079460668542
6.25523044743623 4.36935173479649
6.58140993802512 4.09978178675981
6.8740493489852 3.84688844426225
7.12946097936312 3.61169808492042
7.3503320913801 3.39306711509887
7.54322614667603 3.18739847731712
7.73177677329469 2.98660466097709
7.90777569491641 2.79309023716487
8.06036227596145 2.61007668090759
8.20220246781215 2.43366080571358
8.33125750187086 2.26438824479211
8.4452197904086 2.10289051850582
8.5570955551977 1.9451564476141
8.65161122975686 1.79574295899196
8.73642931192467 1.65226687562193
8.80309472909049 1.51722851211426
8.86588118591431 1.38624253550583
8.91986878040341 1.26073259269904
8.96617975467089 1.1402963939021
9.00337341261919 1.02530966501239
9.02923715281834 0.916385677717483
9.04191239044815 0.814034405485682
9.05048682816827 0.715447602506008
9.04427408327109 0.623808252603017
9.03742703136439 0.53477574916954
9.01913515085121 0.451569336624326
8.98539060480229 0.375358337950093
8.948301351254 0.302427465335561
8.89854991555529 0.235549944747813
8.84677696189201 0.171454888152018
8.78436747041844 0.112703247486231
8.71789449302551 0.0572585531952413
8.65166379158757 0.00376562363581753
8.57162459853465 -0.0435640241084776
8.48853619259967 -0.0880297840973973
8.39134496603003 -0.126323520803602
8.29203649736374 -0.162114343984249
8.19652697022456 -0.197242970013885
8.08861373655832 -0.226843686784439
7.98116619330393 -0.25485253057209
7.86333170855039 -0.278021258260754
7.74016813860672 -0.297915823071098
7.61530296297712 -0.315673334976764
7.48634142162073 -0.330606748084196
7.35236308362571 -0.34247255028556
7.20819378979607 -0.349734349759173
7.05798492450828 -0.353678703984391
6.9040837719224 -0.355048945739803
};
\addplot [semithick, red, mark=x, mark size=3, mark options={solid}, only marks]
table {%
-4.78119115632119 -5.69047807261144
};
\addplot [semithick, sienna1408675]
table {%
-4.78119115632119 -5.69047807261144
-5.30429600928506 -5.32719229100971
-5.77098570990134 -4.98957949559534
-6.18766137402198 -4.67514757391894
-6.56573994968824 -4.3797224944809
-6.90494555397059 -4.10314130240543
-7.20984017061009 -3.84360251553243
-7.47651702344675 -3.60220227661114
-7.69694173072052 -3.3814379569458
-7.90672896134348 -3.16782207013665
-8.09892977369045 -2.96315226551118
-8.2753428022871 -2.76681078794831
-8.44406281547085 -2.57626892851731
-8.59952447134145 -2.39313556887466
-8.73887546727092 -2.21820496649725
-8.86012987510091 -2.05201369798207
-8.97364271321591 -1.89135735083543
-9.07812501621108 -1.73655432164816
-9.16790616695402 -1.58925609729086
-9.24968348951191 -1.44736945826891
-9.31098123569279 -1.3146131430107
-9.36980495647333 -1.18547245153392
-9.42085962454778 -1.06148346238026
-9.45298414325893 -0.945970537090756
-9.48517961720603 -0.833119749871241
-9.49613103959936 -0.729330326556739
-9.50155853234514 -0.629789773153484
-9.49449392698728 -0.536553066190871
-9.4834634212426 -0.446983008316075
-9.45969868299919 -0.363681823627454
-9.4313761073333 -0.284121547603973
-9.39184774121773 -0.210264358856172
-9.35310271540399 -0.138426060341556
-9.29709841862655 -0.074022959539119
-9.24430945703594 -0.0108057204011706
-9.1820126732297 0.0474240753243672
-9.11088138709745 0.100913528722195
-9.0277719588424 0.148751125381701
-8.94458172445255 0.194584934910381
-8.85396000536703 0.23623511925299
-8.75706360163587 0.274093502473409
-8.65353600289087 0.308092999593621
-8.54564343544553 0.338960954081983
-8.43753515895544 0.367997826483116
-8.31394170130539 0.390615035983107
-8.18888645056946 0.411094027835519
-8.05874890579519 0.428377974266872
-7.92371972383048 0.442561212170669
-7.78942732858878 0.455387543583429
-7.65423268115436 0.466401129720133
-7.50642011314116 0.472085196064083
};
\addplot [semithick, red, mark=x, mark size=3, mark options={solid}, only marks]
table {%
-4.54380787075917 -5.92233866647616
};
\addplot [semithick, orchid227119194]
table {%
-4.54380787075917 -5.92233866647616
-5.10100162519041 -5.54260910099382
-5.60388748109227 -5.18805293370509
-6.05576355891173 -4.85723198027972
-6.45975283633365 -4.54878426539126
-6.8188127510146 -4.26141987770287
-7.13574417385978 -3.99391708617868
-7.41319978901704 -3.74511870261618
-7.6536919157833 -3.51392867579135
-7.86642751454538 -3.29639759383666
-8.06385522374051 -3.0871953598724
-8.2464320875033 -2.88610672047586
-8.41461030606661 -2.69291931782805
-8.5688367923472 -2.50742379335626
-8.70955269252857 -2.32941390362644
-8.83719287720572 -2.15868664650184
-8.95218540991726 -1.99504239549752
-9.05495100008255 -1.83828504019375
-9.14590244748526 -1.68822213052551
-9.22544408549685 -1.54466502274143
-9.29397123021505 -1.40742902482324
-9.35186964260216 -1.27633353917634
-9.39951501054807 -1.15120220044317
-9.43727245755425 -1.03186300635355
-9.46549608443951 -0.918148439609068
-9.4845285501102 -0.809895578901289
-9.49470069701895 -0.706946197285141
-9.49633122646198 -0.609146846267648
-9.4897264283402 -0.516348924127538
-9.47517996943841 -0.428408727150876
-9.45297274366646 -0.345187482650602
-9.42337278706134 -0.2665513628316
-9.3866352596772 -0.192371478765848
-9.34300249579732 -0.122523853952252
-9.2927041231954 -0.0568893771508663
-9.23595725145965 0.00464626560079703
-9.17296672868029 0.0621926796629786
-9.10392546509469 0.115854870816729
-9.0290148215936 0.165733374077539
-8.94840506032119 0.211924401534394
-8.86225585396027 0.254520008249074
-8.77071684968615 0.293608275060317
-8.67392828320576 0.329273506958984
-8.57202163777794 0.361596445535384
-8.4651203426411 0.390654493850101
-8.35334050486144 0.41652195194648
-8.23679166826149 0.439270261107716
-8.11557759279951 0.458968254865404
-7.98979704754702 0.475682414690396
-7.85954461025826 0.489477128241619
-7.72491146644108 0.500414948014737
};
\addplot [semithick, red, mark=x, mark size=3, mark options={solid}, only marks]
table {%
5.27116607989859 5.36763109606133
};
\addplot [semithick, gray127]
table {%
5.27116607989859 5.36763109606133
5.75011275340682 5.02494309227316
6.18907108001367 4.70208076032486
6.56419897644157 4.40688216052765
6.89969671221469 4.13080022310348
7.20087207356075 3.87177765503185
7.45710053519472 3.63319468186677
7.68333150807332 3.40971671135282
7.88842936964532 3.19691861922707
8.07945370052798 2.99202173271331
8.26046136930651 2.79371826911017
8.42445355769219 2.60405260837749
8.5794275861441 2.42052870022895
8.71419513495854 2.24646493146617
8.84259552880798 2.07759996321072
8.96017304982663 1.91520947189051
9.06504698905315 1.75979498426425
9.16025258415177 1.61037022706174
9.23442400443662 1.47031722074155
9.30782322719998 1.33343529580821
9.35927026546914 1.20607719572541
9.41044628013719 1.08161592695401
9.44743683092252 0.964201934159397
9.47741592572824 0.851604473522986
9.49404105120016 0.745687437322745
9.50936083778524 0.642742756634326
9.5151790134238 0.5451929739345
9.50693281457577 0.454369418654144
9.49327733207732 0.367599061347713
9.46515769345469 0.287574125842439
9.4307479159699 0.211770676789892
9.38643471623207 0.141233199395475
9.33084252578053 0.0763327531069463
9.27109942568592 0.0148626625571407
9.20116410971286 -0.0413946608475006
9.13711857930708 -0.0973619656587067
9.06289298012086 -0.148223080638393
8.98085996162919 -0.194737800202021
8.88776447524317 -0.235959907920247
8.79821179784482 -0.276358238060765
8.69502062341692 -0.310771942050312
8.58442085200422 -0.341122834380298
8.47604515095843 -0.370371972438686
8.35995880489885 -0.395552883958908
8.23738717607228 -0.41707271650719
8.11635554616697 -0.437402713052088
7.9892272770049 -0.454269898502915
7.84743967465311 -0.465110026936945
7.71006253093437 -0.475733284806293
7.56286132944343 -0.481865264502578
7.41268442106665 -0.485611362411337
};
\addplot [semithick, green01270, mark=*, mark size=3, mark options={solid}, only marks]
table {%
5.21031056537184 -5.96603753689506
};
\addplot [semithick, goldenrod18818934]
table {%
5.21031056537184 -5.96603753689506
4.2242994031561 -5.51882844251806
3.2974068788081 -5.09719186250137
2.42633856449277 -4.6997149308607
1.67988711944863 -4.34851671089884
1.04577888952405 -4.03911466545895
0.508965478250634 -3.76616587810123
0.0563604105068967 -3.52501751583885
-0.323416248733573 -3.31161718108059
-0.640259926823459 -3.12243496413598
-0.902774575892897 -2.95439567058449
-1.11844048849289 -2.80481989628303
-1.29376018751838 -2.67137279542752
-1.44295187936418 -2.5483667598266
-1.57122364322435 -2.43382672960788
-1.68075468521752 -2.32702030650202
-1.77351809697836 -2.22728122885212
-1.85129959231519 -2.13400342527686
-1.91571448416617 -2.04663562411059
-1.9682230723008 -1.96467646493894
-2.0101445963224 -1.88767006355848
-2.04266989405825 -1.81520198624997
-2.06687289224434 -1.74689559340764
-2.08372104440303 -1.68240871635025
-2.09408481986501 -1.62143063458696
-2.09874633790626 -1.56367932395307
-2.09840723186847 -1.50889894889481
-2.09369581982796 -1.4568575747944
-2.0851736508043 -1.40734507860718
-2.07334148859042 -1.36017123825426
-2.05864478898455 -1.3151639831927
-2.0414787204582 -1.27216779038905
-2.02219277305852 -1.23104221156445
-2.00109499557411 -1.19166051907471
-1.9784558966544 -1.15390845914827
-1.95451204162905 -1.11768310243994
-1.92946937319512 -1.08289178297809
-1.9035062808977 -1.0494511175974
-1.87677644139933 -1.01728609886496
-1.84941144889178 -0.986329255333563
-1.82152325262895 -0.956519873697481
-1.79320641643347 -0.927803278090355
-1.76454021313355 -0.900130162356867
-1.73559056520436 -0.873455971655271
-1.70641184140501 -0.847740330211359
-1.67704851790311 -0.822946512450561
-1.64753671125099 -0.799040955087955
-1.6179055896082 -0.775992808059924
-1.58817866778178 -0.753773522439884
-1.55837499096796 -0.732356473697537
-1.52851021151476 -0.711716618839883
};
\addplot [semithick, green01270, mark=*, mark size=3, mark options={solid}, only marks]
table {%
2.60254065376184 5.06573298882138
};
\addplot [semithick, darkturquoise23190207]
table {%
2.60254065376184 5.06573298882138
3.02893910456422 4.76975874755533
3.38236368613298 4.50335927028342
3.69606458436988 4.25537211944649
3.96672639898805 4.02676946349969
4.21365496234258 3.81075121952154
4.41857884986914 3.61334746394276
4.60616725447163 3.4259485136006
4.75649981512583 3.25517150567645
4.89261495388952 3.09298307278092
5.0058227202608 2.94134281190435
5.10437248738055 2.79703432038422
5.19621958611329 2.65758668137982
5.28234620753667 2.52263706906347
5.34794663703708 2.39661424476957
5.41224109192905 2.27364055653466
5.47190972005398 2.15465674946063
5.52033353321734 2.04161041677198
5.55769493861532 1.93439461831084
5.58807330910258 1.83172092805391
5.61156540343013 1.73350594562596
5.62321043331893 1.64120233903956
5.62918659961355 1.55288974219168
5.62405189426266 1.4701679349237
5.60667683799 1.39333580644749
5.58103130670218 1.32114724053352
5.55196634876151 1.25208505007661
5.51555626515497 1.18729087060553
5.46740185672741 1.12805598514067
5.42078027729884 1.07031387390759
5.36285566295928 1.01790756774461
5.30598182243202 0.96705253001959
5.23084133997805 0.923561555272663
5.151610793582 0.883107442792003
5.07194532285378 0.844539642726596
4.98135374470436 0.810997240066512
4.8855522785912 0.780714575780657
4.79134358722274 0.751590278385596
4.68526235995082 0.727666461034827
4.58055242509844 0.704897105648184
4.47342922214584 0.684388642179478
4.35553308344593 0.668641481442053
4.23355028530213 0.655601003203705
4.11670419695817 0.642435356754123
3.99090629600453 0.633376910810011
3.86209849256259 0.626597375874977
3.72685863922769 0.623103700734356
3.59693334365846 0.619304524412408
3.45074479791312 0.621707998413234
3.31216552050761 0.623056391304402
3.15916746499099 0.629993121791321
};
\addplot [semithick, green01270, mark=*, mark size=3, mark options={solid}, only marks]
table {%
5.50926422486168 -1.85014478944775
};
\addplot [semithick, steelblue31119180]
table {%
5.50926422486168 -1.85014478944775
5.18146279491386 -1.7461178949205
4.8701671963013 -1.64699679412334
4.55152482921621 -1.54539502897351
4.24192780047824 -1.44628475451963
3.93131185862077 -1.3466291151145
3.61417393587449 -1.24476298112675
3.30615452247106 -1.14543088891323
3.00425546841668 -1.04773986711078
2.71704303899773 -0.954655096267443
2.44874763188046 -0.8677226032037
2.20723038739578 -0.789269079118471
1.99399304283908 -0.719705871255776
1.8044794871043 -0.657626220192649
1.63334246376363 -0.601391735391429
1.47296130387221 -0.548688329015991
1.32196034070267 -0.499084233152691
1.19176447655821 -0.455990233449988
1.07738857749976 -0.417882147221752
0.964886782661883 -0.380553137540137
0.870106213763179 -0.348750344293029
0.786031393533615 -0.320345118198606
0.705178487360734 -0.293076151520076
0.628831272641135 -0.267320885059341
0.564143050768931 -0.245216705874267
0.507049057642417 -0.225528012266531
0.453587749769672 -0.207054788815178
0.411536291246999 -0.192123450175427
0.361656492845294 -0.174946967106447
0.32113343724996 -0.160686883372144
0.279761989811706 -0.146268783319693
0.25196665402601 -0.136018909842003
0.224611080093802 -0.125979384271103
0.199410466202402 -0.116662334991735
0.180349359947961 -0.10925907078192
0.155455165644061 -0.100169167982362
0.138010638963884 -0.0933823126222051
0.122503622708054 -0.0872351943503193
0.102395728472719 -0.0797625370517181
0.0872799834735993 -0.0738459244723758
0.0790114383057117 -0.0700352675242397
0.0682886221357925 -0.0655313707237279
0.0617121971480458 -0.0623160872897575
0.0544703636690259 -0.0589402593121779
0.0483608291178494 -0.0559429349996472
0.0417848026088001 -0.0528422404380992
0.0364626550145978 -0.0501538797712837
0.0272519234987808 -0.046332691509023
0.0225871452281814 -0.0439124805024318
0.020729421946039 -0.04236585412405
0.0178280960698549 -0.0405337586834659
};
\addplot [semithick, green01270, mark=*, mark size=3, mark options={solid}, only marks]
table {%
-0.809503978690769 5.06218697698483
};
\addplot [semithick, darkorange25512714]
table {%
-0.809503978690769 5.06218697698483
-0.129616364549138 4.7094187255793
0.451059364716547 4.39458573024852
0.945958912976351 4.11289069628911
1.3563777623966 3.86370065225012
1.69536966149368 3.64237409410176
1.96955415632585 3.44648429647575
2.19894372424252 3.269042200031
2.38177088094417 3.11051955039009
2.52057761637563 2.96991855244883
2.63482378717585 2.84047047466538
2.72341522818915 2.72216826274194
2.78612301274751 2.61410932504284
2.84453965528155 2.5097239442081
2.88977896117243 2.41161180727908
2.91390708811361 2.32207940960733
2.9345861320919 2.23574725788945
2.9408344063345 2.15585037272302
2.94230971709464 2.0794213858017
2.93287680998118 2.00824321335092
2.91815984096032 1.94056568645057
2.90111475316714 1.87544663927383
2.87503951721703 1.81484614170843
2.84951256883688 1.75583392690584
2.81597482877071 1.70093300675665
2.78380695894746 1.64727614464552
2.75200370416979 1.59512332025706
2.71581165727765 1.54585913989895
2.67711187514614 1.49887379940429
2.63254285103631 1.45513357969346
2.59678418881512 1.41018939754213
2.555387860071 1.36834784999132
2.51536516636061 1.32746276235925
2.4685254139797 1.2899574630018
2.42797594826948 1.2518575184156
2.37635202583694 1.21834586113681
2.3273015046492 1.18528327707881
2.28081503132286 1.15264346787845
2.23237825040325 1.12175254747889
2.19201689617839 1.08957014608087
2.15060341429759 1.05881377697752
2.11029025420286 1.02880765913537
2.06522211569608 1.00128163179088
2.01960213033122 0.974942289673682
1.97918039572422 0.948037264214729
1.93428233179181 0.923448799185862
1.89441489221258 0.898294515980398
1.85001929208388 0.875423054077701
1.80816741271335 0.852683873912614
1.76194507469984 0.832130817592756
1.71751631300889 0.811887179918956
};
\addplot [semithick, green01270, mark=*, mark size=3, mark options={solid}, only marks]
table {%
-3.95509449913879 -4.77936646051799
};
\addplot [semithick, forestgreen4416044]
table {%
-3.95509449913879 -4.77936646051799
-4.32618370057693 -4.4997906950102
-4.65858031626434 -4.23859783704926
-4.9435234152594 -3.99852887427869
-5.18914949791181 -3.77656503762768
-5.41385914491679 -3.56622309702995
-5.60453767250156 -3.37178614818064
-5.76813523102817 -3.19068856840214
-5.91444531791402 -3.0186072659323
-6.04459555337491 -2.85454159138044
-6.1674445232115 -2.69574244010103
-6.27118678099539 -2.54572254595077
-6.37267912187669 -2.39931249425432
-6.45307953360248 -2.26215754276298
-6.52290417139145 -2.13101806848828
-6.58869485072974 -2.00384923790992
-6.64016019513112 -1.88370825615382
-6.68349813892624 -1.76866234346827
-6.71686711762121 -1.65921238454683
-6.74183365359992 -1.55482871220639
-6.76460839062775 -1.45357185720952
-6.77859459249281 -1.35738296178485
-6.78125043826388 -1.26698206351095
-6.78029260835834 -1.17998044430573
-6.76938537098288 -1.09824582752199
-6.7531716510203 -1.02032325262121
-6.72335966865427 -0.948678651867274
-6.68249899316417 -0.882499153812286
-6.64389804368263 -0.817704913049513
-6.58887591473166 -0.759908954954038
-6.53160100974714 -0.704779361156998
-6.46281067528378 -0.65507868793014
-6.39046637633439 -0.608354840576262
-6.31007156859755 -0.56592774655688
-6.22283987779762 -0.527390997882907
-6.13628750758193 -0.490425103959592
-6.04994614090496 -0.455125640536741
-5.94828681523399 -0.426160331304253
-5.83967075493506 -0.400966699908526
-5.73217606657014 -0.377058870272547
-5.61455556688511 -0.357806793266424
-5.49970083046149 -0.339269643630867
-5.37556450939433 -0.325060084816554
-5.24364170475256 -0.31469657866568
-5.11811335712621 -0.303846802123767
-4.98600614770968 -0.29640266927722
-4.84462631887808 -0.293153216134551
-4.70578600393594 -0.290489377139128
-4.55499985185422 -0.292769031121751
-4.40623281012283 -0.295734598586584
-4.26231566656732 -0.298491204142579
};
\addplot [semithick, green01270, mark=*, mark size=3, mark options={solid}, only marks]
table {%
4.1849124674212 -3.90464512218038
};
\addplot [semithick, crimson2143940]
table {%
4.1849124674212 -3.90464512218038
3.50833405253846 -3.61378459992841
2.88079927401782 -3.34289885366778
2.32333235472725 -3.09852492419189
1.82854946877089 -2.87761439432693
1.41911433533388 -2.68778836152226
1.07506986027072 -2.5220956786886
0.764819282830448 -2.36963344397859
0.502772574663271 -2.23455351224178
0.271175548009888 -2.11034877727092
0.0705914726095148 -1.99711154464041
-0.115418528890463 -1.8898342685577
-0.281513778816298 -1.79006300623283
-0.419347702626273 -1.70024164127091
-0.549896894182255 -1.61400898468527
-0.659176504840868 -1.5355169818818
-0.749541851176045 -1.46400162909678
-0.830138387518658 -1.39666555795405
-0.904024530273307 -1.33254857107906
-0.961917998979239 -1.27439739714311
-1.01693001978427 -1.21823130199736
-1.06400507882729 -1.16553500166565
-1.10225304203605 -1.11653905867975
-1.12950452295303 -1.07185788794213
-1.14812134988416 -1.03074508218928
-1.17116743156587 -0.989246642269377
-1.18388993758572 -0.951768995937617
-1.19326747682732 -0.916183623654379
-1.19787360556145 -0.882891940338245
-1.20113839415652 -0.850837446891209
-1.1971682850791 -0.8217653279317
-1.19006076036072 -0.794417603583619
-1.1885005753642 -0.766164921507476
-1.17593554727422 -0.741959518151261
-1.16674105342292 -0.71745749752805
-1.15418382053025 -0.694663980843724
-1.13586234123164 -0.674277837922583
-1.11896940355797 -0.654117382516251
-1.10462740962252 -0.633830310366031
-1.08331847604824 -0.616258354865493
-1.06618131962012 -0.598036046016657
-1.05627948188459 -0.578232878884207
-1.04391201118945 -0.559751621926415
-1.02463021699414 -0.543909355813044
-1.00165735776444 -0.529717160565992
-0.988577558518071 -0.513081376114554
-0.966024288818329 -0.499809241752374
-0.944402648812828 -0.486755502108779
-0.926381181344953 -0.473107508527669
-0.905142388134927 -0.460901839550613
-0.892599090017854 -0.446548483233732
};
\addplot [semithick, green01270, mark=*, mark size=3, mark options={solid}, only marks]
table {%
4.53989839301795 2.40937239619072
};
\addplot [semithick, mediumpurple148103189]
table {%
4.53989839301795 2.40937239619072
4.58482218925038 2.29922240993052
4.62147672482189 2.19401229832505
4.65011201936482 2.09361349269376
4.6709652260508 1.99790238716107
4.68426092413409 1.90676023797511
4.69021147752735 1.82007304274564
4.68901745703374 1.73773140024471
4.68086812319992 1.65963035161827
4.66594196612608 1.58566920405349
4.64440729797785 1.51575133813173
4.61642289339635 1.44978400027054
4.58213867250254 1.38767808181675
4.5416964207455 1.32934788649579
4.49523053945527 1.2747108880484
4.44286882063392 1.22368747999321
4.38473323925628 1.17620071954141
4.32094075615745 1.13217606775743
4.25160412445888 1.09154112810572
4.17683269243111 1.05422538554856
4.09673319570793 1.02015994836287
4.01141053185518 0.98927729482519
3.92096851045601 0.961511026873647
3.82551057210137 0.936795632793819
3.72514046996821 0.915066260893493
3.61996290802475 0.896258506029015
3.51008413031927 0.880308210725265
3.39561245628099 0.867151282492917
3.27665875748578 0.856723528792158
3.15333687190803 0.848960510922902
3.0257639522899 0.84379741793952
2.9051628188847 0.83754765157247
2.79124291538019 0.830272406686107
2.68347486100746 0.82211195200669
2.58138243853312 0.813190541203981
2.48453766315911 0.803617948993544
2.39255624904663 0.793490881370683
2.30509343727319 0.782894271416328
2.2218401530116 0.77190247086842
2.14251946331362 0.760580346527101
2.06688331011842 0.748984289548525
1.99470949601486 0.737163144771694
1.92579890289207 0.725159066407273
1.85997292593833 0.713008305689553
1.79707110751469 0.700741935445705
1.73694895725786 0.688386515963968
1.67947594637429 0.67596470603854
1.62453366549156 0.66349582262823
1.57201413665063 0.650996352183312
1.521818271068 0.638480416365638
1.47385446518482 0.625960194606412
};
\addplot [semithick, green01270, mark=*, mark size=3, mark options={solid}, only marks]
table {%
2.59246747582498 5.45904872508707
};
\addplot [semithick, sienna1408675]
table {%
2.59246747582498 5.45904872508707
3.09031136111563 5.12765132898499
3.4999365099682 4.8314254783691
3.86862215218408 4.55453678394188
4.19888504635379 4.29588045308387
4.49308978444469 4.05441635310794
4.75345735552338 3.82916537910919
4.98207333975908 3.61920598786548
5.18089575032219 3.42367088979501
5.35176253884948 3.24174389173544
5.49682411649991 3.07247552140521
5.63027231378768 2.9097819250385
5.75237569690441 2.75351693228363
5.86339226559608 2.60353892366623
5.96357025073399 2.4597105665469
6.05314893524519 2.3218985436062
6.13235949169001 2.18997327587568
6.20142582957937 2.06380864240194
6.26056544539943 1.9432816986765
6.30999026825663 1.82827239598977
6.34990749407304 1.71866330387008
6.38052040134984 1.61433933775046
6.40202914167509 1.51518749396529
6.41463149837882 1.42109659411733
6.41852360703173 1.33195704077397
6.41390063184067 1.24766058634896
6.40095739241043 1.16810011690573
6.37988893581396 1.09316945247995
6.35089104943632 1.02276316536459
6.31416071062667 0.956776417631807
6.26989646980174 0.895104818983973
6.21829876428733 0.837644305833189
6.15957016085477 0.784291042306367
6.09391552560086 0.734941343663621
6.02154212052488 0.689491622403298
5.94265962686861 0.647838357109659
5.85748009599684 0.609878083881159
5.76621782930061 0.575507409960682
5.66908918929516 0.544623048976094
5.56631234475343 0.517121876992247
5.4581069533563 0.492901008376122
5.34469378594735 0.47185789028724
5.22629429704516 0.453890414427537
5.10313014678603 0.43889704452054
4.97542267993793 0.426776957840414
4.84339236803884 0.417430198978814
4.70725822106434 0.410757843922854
4.56723717531781 0.406662172421916
4.42354346445816 0.405046846545566
4.27638798073296 0.405817093280168
4.12597763356741 0.408879888978519
};
\addplot [semithick, green01270, mark=*, mark size=3, mark options={solid}, only marks]
table {%
5.17592935558338 -0.807092850715554
};
\addplot [semithick, orchid227119194]
table {%
5.17592935558338 -0.807092850715554
4.94305065548677 -0.755885553351055
4.71012095583368 -0.703884374452305
4.47720423249064 -0.651128700780157
4.2443553812729 -0.597654725835495
4.01162043756554 -0.543495516953697
3.77903686333661 -0.488681103100488
3.54663389914126 -0.433238582714394
3.31443297805722 -0.377192250736396
3.08244819786362 -0.320563743771317
2.85786231011221 -0.26571271774823
2.65316730815524 -0.216735730007254
2.46649053240587 -0.173052807722872
2.29613012249448 -0.134137629773313
2.14053963992402 -0.099512673940818
1.99831415145958 -0.0687448244900747
1.86817763121357 -0.0414413954238683
1.74897155244453 -0.0172465288335301
1.6396445519848 0.00416206848643681
1.53924306105924 0.0230760835704826
1.44690280615281 0.0397592812530579
1.36184109261823 0.0544499410248366
1.28334979196597 0.0673630716220017
1.21078896131966 0.0786924258229214
1.14358103041576 0.0886123357573466
1.08120549783371 0.0972793870500579
1.02319408391569 0.104833948307198
0.969126293118903 0.111401570794556
0.918625343379898 0.117094271639554
0.871354424498456 0.122011712500724
0.827013251599734 0.126242284379425
0.785334883439432 0.129864108088826
0.746082778703471 0.132945958836235
0.709048066545747 0.135548122408981
0.674047010426412 0.137723189574299
0.640918646878386 0.139516794503849
0.609522583158945 0.140968302307898
0.579736939852215 0.142111450107824
0.551456426391609 0.142974945483692
0.524590539181935 0.143583025602077
0.499061873531077 0.143955979854071
0.474804541962104 0.144110638411039
0.451762692678693 0.144060828732656
0.429889123009663 0.143817801735142
0.409143983571416 0.143390629044248
0.389493569668881 0.142786572514815
0.370909197114667 0.142011426993745
0.353366160190486 0.141069837133639
0.336842769912668 0.139965588927369
0.32131947110214 0.138701876527332
0.306778037006477 0.137281544834422
};
\addplot [semithick, green01270, mark=*, mark size=3, mark options={solid}, only marks]
table {%
2.05991675650915 0.206539155446347
};
\addplot [semithick, gray127]
table {%
2.05991675650915 0.206539155446347
1.93992157470653 0.22322249209288
1.83536731944604 0.235651522143036
1.74147304263764 0.245266811504579
1.65576677342949 0.252811050378617
1.56773682011047 0.261436558313911
1.4862732885305 0.268463588539415
1.41704719042139 0.272187747092753
1.34417962926395 0.277376332591657
1.28449989007632 0.278966511397334
1.22813274058789 0.279925515846382
1.17731939461729 0.279574887921961
1.12147970786361 0.28108551662008
1.0739668933561 0.2804365718789
1.02373202642281 0.280942925003717
0.9876470379432 0.277530762360066
0.950816923762602 0.274675243113972
0.907766537702056 0.274009677531849
0.870122017096051 0.272030955716436
0.832335297149751 0.270401252587544
0.799533493483931 0.26757413117674
0.76551544437848 0.265407559122648
0.734566396407751 0.262606860113108
0.71135020939852 0.257768420066605
0.687663980031257 0.253354053153276
0.663254298860555 0.249431943784395
0.637921199799954 0.24605418372275
0.618982824271295 0.241017157269868
0.604478760415551 0.234908919586805
0.581607621545026 0.231568191950456
0.566771430294116 0.226058438003865
0.553043260200581 0.220460001692684
0.532847771124964 0.217040565754779
0.523268563394527 0.210661539951401
0.505945835801231 0.206836632568681
0.491992008341931 0.2022177093476
0.47363697756759 0.199133926911606
0.457444977081339 0.195605646301325
0.448448351639054 0.190119896435925
0.431211754093477 0.187310039807664
0.415824394194593 0.184135136574838
0.40723648304387 0.17910691684282
0.397198447599042 0.174702720487437
0.381176500496568 0.172276436681557
0.376295917100782 0.166678928415058
0.371607852803968 0.16120229607698
0.359674247545392 0.158073781456747
0.345886001686407 0.155663901269184
0.340633820151345 0.150848930679859
0.33775312389401 0.145483228696732
0.324740602351704 0.143316512204919
};
\addplot [semithick, green01270, mark=*, mark size=3, mark options={solid}, only marks]
table {%
-1.33900657467185 -1.99298496561491
};
\addplot [semithick, goldenrod18818934]
table {%
-1.33900657467185 -1.99298496561491
-1.42912392828797 -1.90504209485776
-1.50860416212623 -1.82196799404174
-1.57183539770863 -1.74539181594715
-1.62736606832991 -1.67268972592432
-1.67191594878887 -1.60479728462478
-1.70978987560112 -1.54037213331641
-1.73955712312976 -1.47979855873469
-1.76423052296551 -1.42212731022406
-1.78265304694535 -1.3676645379208
-1.78302678843931 -1.31990891468684
-1.78992988571027 -1.27143385376248
-1.78654850687251 -1.22725862131653
-1.77587831409595 -1.18644220458916
-1.76512965539058 -1.14678322407845
-1.75149071628161 -1.10909575271955
-1.73177420231856 -1.07430438532134
-1.7168636832539 -1.03910947598642
-1.70266396066631 -1.0047176890446
-1.68429314594857 -0.972568066984381
-1.66436212618173 -0.941847171047173
-1.63740803824706 -0.914167461810902
-1.60932856899892 -0.887727868292428
-1.5877693055525 -0.860209799523075
-1.56490291935823 -0.833945643501473
-1.54087491812278 -0.808867978341975
-1.51149793298126 -0.786210171625704
-1.4783497326693 -0.765472111285298
-1.4497005890795 -0.744148015365566
-1.4220448164608 -0.723274497846966
-1.39207011648949 -0.703826923374393
-1.35926438287056 -0.685937554499951
-1.32960811755349 -0.667790661698532
-1.30741371566679 -0.648077742729412
-1.2782006501773 -0.631133693311611
-1.25371986778266 -0.613408304229489
-1.22693366429119 -0.597001424599467
-1.20537773964784 -0.579633384476949
-1.1829095278918 -0.563136764180294
-1.16231926789487 -0.546657901372197
-1.13194192811056 -0.533683553749585
-1.1117797410455 -0.518187435634471
-1.085363261676 -0.505107407688143
-1.06081235753739 -0.491986717320498
-1.04065428759649 -0.478055520710213
-1.01916440599663 -0.465022758576883
-1.00005849233926 -0.451759210495655
-0.977941510463848 -0.439873445353402
-0.955492453224785 -0.428546218619523
-0.926225674305484 -0.419711181456702
-0.899590147676451 -0.410514636408685
};
\addplot [semithick, green01270, mark=*, mark size=3, mark options={solid}, only marks]
table {%
-2.50964808486869 1.22643416559638
};
\addplot [semithick, darkturquoise23190207]
table {%
-2.50964808486869 1.22643416559638
-2.23995324459091 1.13060849170334
-1.98879388093751 1.04101051917003
-1.7668176312261 0.960798444742914
-1.56554074189176 0.887380298319504
-1.38460912571815 0.820613281660994
-1.21934499015281 0.759058643956392
-1.06982786660423 0.70271075466766
-0.935880296566543 0.651487561858028
-0.81674642573053 0.60513440639649
-0.709941783203049 0.562879487082457
-0.6132732468948 0.524041670703141
-0.520731942873619 0.48679782641449
-0.444826822570884 0.454887581243355
-0.369485925844746 0.423462674040099
-0.302025189220797 0.394709349128895
-0.250129817590401 0.370917315291349
-0.205488775871471 0.3495693054416
-0.158595625191501 0.327798589579929
-0.123093101953643 0.309694232210787
-0.0924739716893574 0.293285587154338
-0.0639740837530727 0.277732267287887
-0.0297525232468318 0.260673814012072
-0.00107674767692615 0.245491283773153
0.0152397444707725 0.234217731188989
0.0343718233445518 0.222282151270547
0.0466011313923698 0.212598561296848
0.0673578986732991 0.200525492330923
0.0839219905578123 0.18988386138531
0.0971631491649574 0.180403870081887
0.112973113296196 0.170310066468321
0.117242242168742 0.163834429810119
0.126621496270831 0.155965096630468
0.136669457165804 0.148036304873624
0.140141957439752 0.142218587805731
0.14528085059445 0.136028700731647
0.143025239801272 0.132183640512028
0.142163281362359 0.128035717366839
0.140670382267805 0.124190824348396
0.141194790994307 0.119850780436691
0.142244969286384 0.115462321583416
0.141824547352588 0.111622130910093
0.144665414592491 0.106906165637375
0.143770942087512 0.103414367006374
0.144531727740221 0.0995228096236676
0.140385188453617 0.0971996980842511
0.138396857333153 0.0943175459073326
0.142299498424318 0.0897563155622673
0.138922717907755 0.0874713918479511
0.131969764839274 0.0863417456595206
0.129910264182226 0.0838203452976021
};
\addplot [semithick, green01270, mark=*, mark size=3, mark options={solid}, only marks]
table {%
2.22801544516873 -1.91135733917186
};
\addplot [semithick, steelblue31119180]
table {%
2.22801544516873 -1.91135733917186
1.9118313277359 -1.7818612326721
1.62844100204313 -1.66343184663223
1.37471234707713 -1.55506362470997
1.14780113164526 -1.45584233560775
0.945124215541488 -1.36493660632142
0.764335313020402 -1.28159026294583
0.603303077636318 -1.20511540308989
0.460091289416207 -1.13488613087203
0.332940945224689 -1.07033289174698
0.220254071240874 -1.01093735011781
0.120579092876472 -0.956227757868062
0.0325976123813538 -0.905774766658929
-0.0448875420494183 -0.859187641119214
-0.112963119504166 -0.816110833954539
-0.172615214508407 -0.776220887551743
-0.224738102441687 -0.739223629889574
-0.270142235203348 -0.704851635516965
-0.309561481575357 -0.672861925052743
-0.343659688929069 -0.643033879119961
-0.373036635788624 -0.615167344876113
-0.398233438234005 -0.589080915356972
-0.419737467145688 -0.564610363734457
-0.437986827810003 -0.541607216313587
-0.453374448374986 -0.519937449674443
-0.466251819030429 -0.499480298814893
-0.47693241954728 -0.480127164479883
-0.485694868918487 -0.461780609083667
-0.492785827267433 -0.444353431751301
-0.498422676905865 -0.427767814033319
-0.502796006408161 -0.411954528789778
-0.506071918802841 -0.396852205603328
-0.508394182447459 -0.382406646871343
-0.5098862408336 -0.368570189449495
-0.510653095450234 -0.355301107378131
-0.510783073903595 -0.342563051822507
-0.510349493738394 -0.330324524902004
-0.509412230818316 -0.318558384575376
-0.508019199693905 -0.307241378191737
-0.506207752104447 -0.296353702713434
-0.50400599861942 -0.285878589969484
-0.501434057416979 -0.275801915609562
-0.498505233314802 -0.266111830700724
-0.495227129405766 -0.256798415144331
-0.491602693000923 -0.247853352291154
-0.487631197038883 -0.239269624300324
-0.483309157677824 -0.231041227924599
-0.478631188437985 -0.223162910512425
-0.473590791002533 -0.21562992609814
-0.468181082607251 -0.208437811507614
-0.462395459848411 -0.20158218243922
};
\addplot [semithick, green01270, mark=*, mark size=3, mark options={solid}, only marks]
table {%
2.03053050095627 0.819732303586688
};
\addplot [semithick, darkorange25512714]
table {%
2.03053050095627 0.819732303586688
1.97170820817513 0.802311179338065
1.91576974915755 0.784876207054775
1.86251998526892 0.767466769541867
1.81177394060545 0.75011949731657
1.76335601577358 0.732868507280331
1.71709933364678 0.715745600908276
1.67284520307094 0.698780426324655
1.63044268721823 0.682000608395461
1.58974826397366 0.665431850746785
1.55062556638726 0.649098013406496
1.51294519184749 0.633021169564943
1.47658456923458 0.617221644755262
1.44142787390301 0.601718041563588
1.40736598092538 0.586527252792393
1.37429644760951 0.571664465814965
1.34212351688094 0.557143160674641
1.31075813370556 0.542975104298039
1.28011796731474 0.529170343006665
1.25012743258824 0.515737195325509
1.22071770454889 0.50268224690062
1.19182672052756 0.490010349150095
1.16339916516589 0.477724623084825
1.13538643403606 0.465826469547024
1.10774657227079 0.454315586926677
1.08044418520836 0.443189997229167
1.05345031866673 0.432446081182314
1.02674230706209 0.422078622888729
1.00030358817962 0.41208086435068
0.974123483983092 0.402444570020517
0.948196947411905 0.393160101361219
0.922524275656692 0.38421650123968
0.897110790922797 0.375601587821091
0.87196649018258 0.367302057487097
0.847105665878515 0.359303596164334
0.822546499966542 0.351590998324344
0.798310634080154 0.344148292801607
0.774422718947364 0.336958874474265
0.750909946502755 0.330005640762743
0.727801568402953 0.323271131825485
0.705128404874419 0.316737673268889
0.68292234799599 0.310387520140626
0.661215863644089 0.304203000942109
0.640041496405353 0.298166660377099
0.619431381789486 0.292261399549218
0.599416770054629 0.286470612331449
0.580027565889136 0.280778316655255
0.561291888078583 0.275169279505288
0.543235653126488 0.269629134457398
0.525882186593758 0.264144490662075
0.509251865677478 0.258703032251834
};
\addplot [semithick, green01270, mark=*, mark size=3, mark options={solid}, only marks]
table {%
0.217118631662959 2.18368891186708
};
\addplot [semithick, forestgreen4416044]
table {%
0.217118631662959 2.18368891186708
0.398478683197979 2.0710058075072
0.558555667084775 1.96641135307487
0.699453356560306 1.86920805681823
0.823070034039298 1.77876397929486
0.931117861318636 1.69450660792295
1.02514049960282 1.61591728482144
1.10652914102835 1.54252613688875
1.17653709769204 1.47390746200564
1.23629308000143 1.4096755297148
1.28681328333205 1.34948075877137
1.32901239037421 1.29300623761182
1.36371358607262 1.23996455708903
1.39165767260387 1.19009492780042
1.41351136330943 1.14316055702207
1.42987482682141 1.09894626268177
1.44128854570963 1.05725630398169
1.44823954777118 1.01791241023816
1.45116706251508 0.980751991263029
1.45046765040631 0.945626514185614
1.44649984797158 0.91240003302357
1.43958836788569 0.880947858570683
1.43002788960652 0.851155357293333
1.41808647296706 0.822916868928686
1.40400862432727 0.796134733368222
1.3880180424001 0.770718418201152
1.37032006866414 0.746583738993839
1.35110386532863 0.723652165002747
1.33054434209807 0.701850203568541
1.30880385146775 0.681108856925556
1.28603367094435 0.661363145591406
1.26237528940606 0.642551692882743
1.23796151377361 0.624616365441345
1.21291741123987 0.607501964955539
1.18736110148308 0.591155966530675
1.16140441255291 0.575528299403823
1.13515341345445 0.560571165916338
1.10870883585038 0.546238894857505
1.08216639674467 0.53248782547665
1.05561703349075 0.519276218633111
1.02914706197496 0.506564191716233
1.00283826835346 0.494313674123474
0.976767944259869 0.48248838023621
0.951008874947076 0.471053796981568
0.925629289372448 0.4599771832164
0.900692780778892 0.449227578317589
0.876258205859643 0.438775817512312
0.85237957012071 0.428594551633509
0.829105906569073 0.418658269140222
0.806481154356196 0.408943318399967
0.784544043494493 0.399427928391129
};
\addplot [semithick, green01270, mark=*, mark size=3, mark options={solid}, only marks]
table {%
0 0
};
\addplot [semithick, crimson2143940]
table {%
0 0
0 0
0.000613505815992204 -0.000184074123263327
0.00177796219085373 -0.000532819663686005
0.003430555241633 -0.00102683743386827
0.00550835086872815 -0.00164677145026955
0.00794848001887223 -0.00237336258114413
0.0106883646896778 -0.00318751427742568
0.0136659785541698 -0.00407036851144144
0.0168201360984494 -0.00500339006007508
0.0200908042076983 -0.00596845728873749
0.0234194302078807 -0.006947957621041
0.0267492804748474 -0.00792488591701692
0.0300257838606466 -0.00888294403050605
0.0331968743596448 -0.00980663987424673
0.0362133276449747 -0.0106813843892762
0.0390290863486811 -0.011493584893477
0.041601569236072 -0.012230733372207
0.0438919597349924 -0.0128814883715709
0.045865469622377 -0.0134357492614989
0.0474915740413763 -0.0138847217507363
0.0487442144200726 -0.0142209736583495
0.0496019662843999 -0.0144384800755166
0.0500481694001079 -0.014532657186228
0.0500710181379355 -0.0145003841549877
0.0496636104287755 -0.0143400126325477
0.0488239541575393 -0.0140513635759232
0.0475549303314897 -0.0136357112251751
0.0458642128467408 -0.0130957542254464
0.0437641451610864 -0.0124355740271963
0.0412715746579559 -0.01166058083922
0.0384076459507976 -0.0107774475465731
0.035197554825307 -0.00979403213771732
0.0316702649445398 -0.00871928931083838
0.0278581898451388 -0.00756317204722093
0.0237968431279192 -0.00633652404871141
0.0195244600894403 -0.00505096403564976
0.0150815943497358 -0.00371876299029426
0.0105106933022416 -0.0023527155078819
0.0058556564426292 -0.000966006482354598
0.00116138082160999 0.00042792559415222
-0.00352670198889386 0.00181552740018537
-0.00816309292781013 0.00318327027919255
-0.0127026919339711 0.00451778910475224
-0.0171012593446673 0.0058060191650841
-0.0213158676964748 0.00703532973155971
-0.025305339507426 0.00819365300603833
-0.0290306667509527 0.00926960718527963
-0.0324554079070697 0.0102526124369769
-0.0355460586931456 0.0111329986505225
-0.038272392833106 0.0119021039057542
};
\end{axis}

\end{tikzpicture}

%% file: fig/ltiRnn-phase-space.tex
\begin{tikzpicture}

\definecolor{crimson2143940}{RGB}{214,39,40}
\definecolor{darkgray176}{RGB}{176,176,176}
\definecolor{darkorange25512714}{RGB}{255,127,14}
\definecolor{darkturquoise23190207}{RGB}{23,190,207}
\definecolor{forestgreen4416044}{RGB}{44,160,44}
\definecolor{goldenrod18818934}{RGB}{188,189,34}
\definecolor{gray127}{RGB}{127,127,127}
\definecolor{green01270}{RGB}{0,127,0}
\definecolor{mediumpurple148103189}{RGB}{148,103,189}
\definecolor{orchid227119194}{RGB}{227,119,194}
\definecolor{sienna1408675}{RGB}{140,86,75}
\definecolor{steelblue31119180}{RGB}{31,119,180}

\begin{axis}[
    width=\textwidth,
    height=5cm,
    yticklabels={,,},
    xticklabels={,,},
tick align=inside,
tick pos=left,
x grid style={darkgray176},
xlabel={$x_1$},
xmin=-15, xmax=15,
xtick style={color=black},
y grid style={darkgray176},
xticklabels={,-10,0,10},
ymin=-10, ymax=10,
ytick style={color=black}
]
\addplot [semithick, red, mark=x, mark size=3, mark options={solid}, only marks]
table {%
4.37444990644978 5.38906124561606
};
\addplot [semithick, steelblue31119180]
table {%
4.37444990644978 5.38906124561606
4.84655576187762 4.81840346371794
5.25126623308329 4.27700248519418
5.59221154429215 3.76523508405172
5.9023692922509 3.28944889447894
6.19808337602267 2.85040817184777
6.48150821495944 2.44511763809645
6.75467377929317 2.07078760488078
7.01949590328111 1.7248191570629
7.27778641049219 1.40479054642089
7.53126313453604 1.1084446844667
7.78155991260881 0.833677630153169
8.03023662365087 0.578527974755028
8.27878933769995 0.341167032359655
8.52866063820353 0.119889750235222
8.78125017465231 -0.0868937411162058
9.03792549894153 -0.280666013140756
9.30003323538926 -0.462809729545675
9.56891063137322 -0.634614606846008
9.84589753312344 -0.797284005674322
10.1323488293632 -0.951941209018854
10.4296474042594 -1.09963543939503
10.7392176405622 -1.24134766299463
11.0625395139219 -1.37799622509488
11.4011633201969 -1.51044235744789
11.7567250791609 -1.63949559501195
12.1309626604016 -1.7659191362356
12.5257326804242 -1.89043517816962
12.9430282240647 -2.01373025496924
13.3849974483176 -2.13646060586782
13.8539631326231 -2.25925759646438
14.3524432465873 -2.38273321518139
14.8831726140521 -2.50748566502752
15.4491257614381 -2.63410506935498
16.0535410483902 -2.76317930914352
16.6999461900024 -2.89530000848758
17.3921852923391 -3.03106868441903
18.1344475366392 -3.17110307698053
18.931297662555 -3.31604367558359
19.7877084170781 -3.46656045815437
20.7090951535172 -3.62335986040013
21.7013527840725 -3.78719199373175
22.7708953102781 -3.95885813196447
23.924698177938 -4.13921848890042
25.1703437272536 -4.3292003112845
26.516070034731 -4.5298063144303
27.9708234712797 -4.74212349104616
29.5443153307912 -4.96733232746506
31.2470829155583 -5.20671646560697
33.0905554993188 -5.46167285359224
35.0871256256463 -5.7337224329913
};
\addplot [semithick, green01270, mark=*, mark size=3, mark options={solid}, only marks]
table {%
-5.07318404917714 4.85075096054016
};
\addplot [semithick, darkorange25512714]
table {%
-5.07318404917714 4.85075096054016
-4.6752374468116 4.72592518707322
-4.31166155591753 4.57923499401042
-3.89710142610678 4.45604969419986
-3.49549551319691 4.31707963502487
-3.11206899077837 4.17458270434625
-2.7474846662984 4.03064392255252
-2.40145918529663 3.88509899966439
-2.0709534924483 3.72952319978296
-1.76080064979955 3.57661318795118
-1.4697080620157 3.42432913588913
-1.19871215165238 3.27697342548724
-0.945896268496143 3.13120988320254
-0.709610679881105 2.98354188529938
-0.491798476302183 2.84018662530723
-0.291116637331392 2.69926876590184
-0.105981716012379 2.55764572487615
0.0609720511397814 2.42384250053055
0.213015150389525 2.29112208888198
0.34802593704277 2.16690312880191
0.470593808913126 2.04063220615542
0.575335157184878 1.92850353718336
0.668826563639978 1.81552732258538
0.748228281330699 1.71056556998664
0.816214120108499 1.60847980324056
0.870096494697759 1.51863542262099
0.913086344752014 1.43537100260629
0.948071388736035 1.35236896695301
0.973699418424009 1.27416553557276
0.99006956280028 1.20220561583566
0.997427558383122 1.13785973285953
0.996459692396281 1.08152198720217
0.991673632553561 1.02227347847385
0.978133813821924 0.974277523559285
0.960746830017765 0.926276741018842
0.938891754671889 0.880331777428583
0.911314853990151 0.840945581156125
0.880370524369138 0.803095707106806
0.842131275340298 0.779050869841611
0.803862871142012 0.751099390445698
0.76348152500722 0.724452021172158
0.724738177700662 0.704956119075193
0.688387381794886 0.681679630551335
0.653921016232041 0.663651705236289
0.621493193788324 0.644731827838607
0.590798119614127 0.629793059611641
0.561939666332165 0.613953023719682
0.534908649250348 0.593225983238925
0.509107552268615 0.580064673619729
0.48487298862612 0.565134882367437
0.46194902301474 0.552298967986634
};
\addplot [semithick, red, mark=x, mark size=3, mark options={solid}, only marks]
table {%
-4.63963370552342 -5.45600729794191
};
\addplot [semithick, forestgreen4416044]
table {%
-4.63963370552342 -5.45600729794191
-5.11825636958106 -4.86825905939341
-5.52793387193186 -4.30967196502585
-5.87228781438333 -3.78071015107604
-6.19004393573813 -3.28902348777728
-6.48581255240655 -2.83263182512088
-6.76115009609902 -2.40894292587902
-7.017535204417 -2.01557319871801
-7.25638044488984 -1.6503302298122
-7.4790431158891 -1.31119648482948
-7.68683517272476 -0.99631410575211
-7.88103232713826 -0.703970733775267
-8.06288236876055 -0.432586295614213
-8.23361275784872 -0.180700695990626
-8.39443753969918 0.0530376361012426
-8.54656363251238 0.269882395488917
-8.691196542105 0.470999287691922
-8.82954555868657 0.657475195756214
-8.96282849289755 0.83032666547676
-9.09227601040443 0.990507745096106
-9.21913562653056 1.13891721362536
-9.34467542463523 1.27640523035447
-9.47018756421102 1.40377943686892
-9.59699164692426 1.52181054193891
-9.72643801105727 1.63123741896398
-9.85991102700596 1.73277174521192
-9.99883246863251 1.82710221185666
-10.1446650373631 1.91489833376829
-10.2989161179536 1.99681388811339
-10.4631418468256 2.07349001105997
-10.6389515758099 2.14555798222462
-10.8280128160403 2.21364172692853
-11.0320567486313 2.27836006682165
-11.2528843906884 2.34032874997271
-11.4923735071469 2.40016229208865
-11.7524863609775 2.45847566110566
-12.0352783964562 2.51588583797083
-12.3429079525255 2.57301328699809
-12.6776471058362 2.63048336972456
-13.0418937458957 2.68892773670625
-13.4381849879368 2.7489857321714
-13.8692120327288 2.81130584689175
-14.3378365866477 2.87654725503748
-14.8471089599895 2.94538147115207
-15.4002879668441 3.01849416372379
-16.0008627559166 3.09658716214777
-16.6525767086103 3.18038069417615
-17.3594535485545 3.27061589125566
-18.1258258156838 3.36805759946631
-18.9563658680729 3.47349753411802
-19.8561195861105 3.58775781645382
};
\addplot [semithick, green01270, mark=*, mark size=3, mark options={solid}, only marks]
table {%
5.78886904899375 -4.25151721984619
};
\addplot [semithick, crimson2143940]
table {%
5.78886904899375 -4.25151721984619
5.48186335937119 -4.13841990005301
5.16835874282159 -4.03243509570579
4.8480737079722 -3.93349275611871
4.52068016990867 -3.84154256164785
4.18580253478408 -3.75655392745183
3.84484815328716 -3.67813086360189
3.52785618882734 -3.59985829136921
3.22506350556115 -3.51674792722803
2.93289364075403 -3.42731060718886
2.6520393697233 -3.33258886016818
2.38307287744107 -3.23356988735211
2.12645242209352 -3.13118463050761
1.88252906418726 -3.02630716831552
1.65155342526935 -2.91975442083625
1.43368244285831 -2.81228614269928
1.22898609063014 -2.70460518609899
1.03745403526299 -2.59735801517378
0.859002203617095 -2.49113545383685
0.69347923611198 -2.38647364961138
0.540672804263033 -2.28385523649509
0.400315772354743 -2.18371068033751
0.2720921851612 -2.08641979065533
0.155643065478097 -1.99231338323579
0.0505720070073845 -1.90167507828599
-0.0435494501609291 -1.81474321927595
-0.1271766725924 -1.73171289799895
-0.200787024633922 -1.65273807173358
-0.264875205676486 -1.57793375874171
-0.319948840168671 -1.50737829867804
-0.366524325611483 -1.44111566482185
-0.40512294243724 -1.37915781537488
-0.436267228400174 -1.32148707140346
-0.460477618882128 -1.2680585093415
-0.478269353341511 -1.2188023563181
-0.490149647006072 -1.17362637693221
-0.496615125829088 -1.13241824047001
-0.498149521693674 -1.09504785795252
-0.495221623860667 -1.06136967881354
-0.488283481712241 -1.03122493744355
-0.477768852946271 -1.00444384029693
-0.464091890526437 -0.980847684748367
-0.449431826106458 -0.959134235441656
-0.435376152862445 -0.938287473395576
-0.421900554386941 -0.918238742971758
-0.408981261422603 -0.898918871294508
-0.396594991762525 -0.880258651167209
-0.384718901604026 -0.86218933061966
-0.373330547861535 -0.844643103976332
-0.362407860896235 -0.827553599329508
-0.351929127073704 -0.810856357345963
};
\addplot [semithick, red, mark=x, mark size=3, mark options={solid}, only marks]
table {%
4.99558976766198 5.30797448402492
};
\addplot [semithick, mediumpurple148103189]
table {%
4.99558976766198 5.30797448402492
5.45463571645278 4.72108760433917
5.84622245873534 4.1602444035601
6.21488574617279 3.64171125603995
6.54464306165209 3.15203193340488
6.87576792951321 2.70607530284674
7.19193725511201 2.29154079956319
7.49613647834247 1.90648560528779
7.758674055592 1.53586563283483
8.05908470555262 1.21240092132075
8.36274875041723 0.913761969097723
8.61597500431237 0.614909126543142
8.87551662669215 0.341800793537528
9.13255049862774 0.0871780906096672
9.37530830338138 -0.155934349828909
9.66720198711714 -0.3625968757479
9.93832786822842 -0.568662503014759
10.2235081731701 -0.75891443799082
10.4815726640784 -0.952837727029202
10.7795986695033 -1.12127966812727
11.097856850249 -1.27789945631774
11.4436588146 -1.42250031090025
11.8122024970416 -1.56023637536889
12.1938176308648 -1.69726840379041
12.5790397102361 -1.83829388888152
13.012067845562 -1.96520123408092
13.4439391942044 -2.10244248207123
13.9432520563236 -2.22096532134778
14.4629033259941 -2.34625499381703
14.9826907028919 -2.48681382776573
15.5606107786547 -2.61730456822789
16.1553119540602 -2.75921474462556
16.8195097156254 -2.89084056856899
17.5165794127355 -3.03248063228686
18.2805585871008 -3.17111559570756
19.1388394863546 -3.30014561704541
20.0298975279306 -3.45132069930141
21.0075981906991 -3.6020594287506
22.0238010690768 -3.7770707389274
23.1389109045861 -3.95120993820666
24.3914405733581 -4.11505823652918
25.7134514155659 -4.30565155071913
27.1726367585569 -4.49646532509764
28.7268570608821 -4.71276635273712
30.4071844737931 -4.9445010835139
32.2416335231773 -5.18602728749583
34.2312940261261 -5.44497992624897
36.3839770836241 -5.72536627021622
38.6864441133997 -6.03963441597044
41.1697420569484 -6.37997520055974
43.8597306041838 -6.74355819736493
};
\addplot [semithick, red, mark=x, mark size=3, mark options={solid}, only marks]
table {%
-4.78119115632119 -5.69047807261144
};
\addplot [semithick, sienna1408675]
table {%
-4.78119115632119 -5.69047807261144
-5.28600780982965 -5.07654038367932
-5.7196903945444 -4.49016397519597
-6.08503483856867 -3.93398533826257
-6.44425790833068 -3.42635774743775
-6.79917937602813 -2.96069732478241
-7.16092906648656 -2.53633865584514
-7.51252505289169 -2.14069701009637
-7.81906951923792 -1.75661809736551
-8.14753627831794 -1.41233489229148
-8.45793432847834 -1.08526830613318
-8.75725701556442 -0.777111500207714
-9.08508916877929 -0.502052107739349
-9.41775869016544 -0.244553366030268
-9.7413082153707 0.00356811526537637
-10.0439381104114 0.24772513760366
-10.37423346675 0.46757865736692
-10.7291971566642 0.669146891291948
-11.0829009724581 0.866733097935803
-11.4675996986861 1.04777369443961
-11.823650379816 1.24032010326529
-12.2359242555577 1.40694436723451
-12.6844566649349 1.56244253427474
-13.1162742951373 1.73179770881917
-13.6222922641126 1.87556558496231
-14.1037990547714 2.04132821993509
-14.6336367252081 2.19540417037831
-15.1812986157655 2.35452467214326
-15.7894695598731 2.50191423717816
-16.4198529109691 2.65818745254839
-17.1134568925499 2.8070153688864
-17.8425187056744 2.96437834601161
-18.6686223556683 3.10650143797363
-19.5108985964649 3.27382448447294
-20.4699963556064 3.42350941100035
-21.4940558596297 3.5858109107863
-22.5903072952933 3.76012320809258
-23.7475734165441 3.95377605687465
-25.0284759601593 4.14273104379834
-26.4075527212997 4.34534701532508
-27.8976967772582 4.56061305940392
-29.5055802198557 4.79093717525682
-31.2514832453304 5.03333608242517
-33.167486504539 5.28168347710717
-35.1918092782552 5.57041883530894
-37.400287273522 5.87032919527632
-39.7905799519526 6.19207323783035
-42.3781626370924 6.53761236278778
-45.2071856722121 6.89803464967055
-48.2905421633276 7.28085085043512
-51.5909094317525 7.71362444934016
};
\addplot [semithick, red, mark=x, mark size=3, mark options={solid}, only marks]
table {%
-4.54380787075917 -5.92233866647616
};
\addplot [semithick, orchid227119194]
table {%
-4.54380787075917 -5.92233866647616
-5.07609508418988 -5.29878252756243
-5.53511934092671 -4.70515330768462
-5.9246080742688 -4.1420345766348
-6.28369764074988 -3.61724219943294
-6.62147166167855 -3.12978091040003
-6.93975594233378 -2.67686376666187
-7.24030498615533 -2.25592387807334
-7.52481549158703 -1.86459578905959
-7.79493898173969 -1.50069809425152
-8.05229362695313 -1.16221720555757
-8.29847532070095 -0.847292195441482
-8.53506807016749 -0.554200647572706
-8.76365376418576 -0.281345451725348
-8.98582138301179 -0.0272424848641949
-9.20317571657926 0.209490875187892
-9.41734566038838 0.430146453219789
-9.62999216099499 0.63593524360199
-9.84281588614512 0.827996684259187
-10.0575646979146 1.00740727874852
-10.2760410107357 1.17518860557334
-10.5001091198987 1.33231475228291
-10.7317025899883 1.47971921069047
-10.972831796736 1.61830126865521
-11.2255917199363 1.74893193328687
-11.4921700893774 1.87245942011756
-11.7748559901818 1.98971424271469
-12.0760490385501 2.10151393735686
-12.3982692436593 2.20866745773481
-12.7441676764143 2.31197927514947
-13.1165380709163 2.41225322033624
-13.5183294899172 2.51029610382896
-13.952660191242 2.60692115267009
-14.4228328382011 2.70295130225848
-14.9323512034705 2.79922238318959
-15.4849385228265 2.89658624407072
-16.0845576625835 2.99591385247854
-16.7354332726571 3.09809841745776
-17.4420761059687 3.20405857823524
-18.2093096945125 3.31474170513945
-19.0422995829281 3.43112736007217
-19.9465853319867 3.55423096528035
-20.928115517128 3.68510773062738
-21.9932859612097 3.82485689107343
-23.1489814561161 3.97462630765667
-24.4026212449512 4.13561748693515
-25.7622085554011 4.30909107562255
-27.2363844956644 4.49637288904973
-28.8344866473052 4.69886053413464
-30.5666127146933 4.91803068977176
-32.4436896185755 5.15544710999136
};
\addplot [semithick, red, mark=x, mark size=3, mark options={solid}, only marks]
table {%
5.27116607989859 5.36763109606133
};
\addplot [semithick, gray127]
table {%
5.27116607989859 5.36763109606133
5.73682829777706 4.76092665571896
6.18324515282319 4.20289697293174
6.5531359712653 3.65894079286886
6.91345259615533 3.15886522338348
7.27752458229437 2.7029641500379
7.60461983953594 2.26889567294554
7.93677981293931 1.87302982243289
8.25337388836579 1.50181770675158
8.56589413229063 1.15765505503832
8.8951013722654 0.845668351392904
9.2087854860347 0.547714509000023
9.5451234925202 0.278312098359642
9.85113248532372 0.0110511799763238
10.1917499636286 -0.226332994450853
10.5492719618957 -0.446925794808571
10.9158605968881 -0.657075903925521
11.3070034504887 -0.852436771502556
11.6677722871041 -1.05854279383201
12.0942533807572 -1.23343915688493
12.4880241043925 -1.42480231074848
12.9515475519241 -1.58722777001125
13.4231803292969 -1.7535193474003
13.9378420460484 -1.9095948720295
14.4674275064177 -2.07070678205782
15.0719903286354 -2.21290417658094
15.7182379861992 -2.35658755858962
16.3857402968188 -2.51232532173555
17.1174489896539 -2.66253823999628
17.873352257569 -2.82795587856876
18.6939876521432 -2.99174392169458
19.5664034658336 -3.16325447540038
20.4866587424122 -3.34621705096354
21.4928264749269 -3.52670709972771
22.5619845017068 -3.71948731088492
23.7786966597901 -3.89270407544222
25.0775286488927 -4.08419362284352
26.4753743463432 -4.28976512420708
27.9629410784151 -4.51734653859963
29.6190373749084 -4.73776995840994
31.3734501945037 -4.9904686830856
33.2626400881428 -5.26254070679603
35.3451560776761 -5.53605360326825
37.58880044555 -5.83551133164972
40.0117490920267 -6.16000340399229
42.6682768421889 -6.49528765695627
45.5407387137489 -6.86123500269193
48.6029836221335 -7.27751336685044
51.9631933738674 -7.70552514327479
55.5767709326874 -8.17978388411555
59.4971111057122 -8.6881534507808
};
\addplot [semithick, green01270, mark=*, mark size=3, mark options={solid}, only marks]
table {%
5.21031056537184 -5.96603753689506
};
\addplot [semithick, goldenrod18818934]
table {%
5.21031056537184 -5.96603753689506
4.74550107849869 -5.73264263924639
4.26274321268524 -5.5084216580958
3.75934539733893 -5.29218906894302
3.2630477863801 -5.07638394086606
2.79330611567913 -4.857186321271
2.35040316773211 -4.63604916713105
1.93445652426171 -4.41431895377414
1.54543132690565 -4.1932374046917
1.18315276773217 -3.97394356545988
0.847318272645869 -3.75747618435487
0.537509345348383 -3.54477636437306
0.253203043843644 -3.33669045346906
-0.00621693447210273 -3.13397314188699
-0.241449579313649 -2.93729073746919
-0.453266322264047 -2.74722459177236
-0.642501233184648 -2.56427465169336
-0.810041668663153 -2.38886311309992
-0.956819380109316 -2.22133815466809
-1.08380208656026 -2.06197773174484
-1.19198551492657 -1.91099341157841
-1.28238590828498 -1.76853423268767
-1.35603300089139 -1.63469057247718
-1.41396345683791 -1.50949800844536
-1.45721476769666 -1.39294115948253
-1.4868196030696 -1.28495749481671
-1.50380060668634 -1.18544109914019
-1.50916562954814 -1.09424638334596
-1.50390339059641 -1.01119173112259
-1.48897955447579 -0.936063072406928
-1.46533321515684 -0.868617375380926
-1.43387377347082 -0.808586049328407
-1.39547819598067 -0.755678251246872
-1.35098864205991 -0.70958408964429
-1.30121044556765 -0.669977719448506
-1.24691043708564 -0.636520322423596
-1.18881559231798 -0.608862967929726
-1.12761199193851 -0.586649349286804
-1.06394407790259 -0.56951839141327
-1.00258348921378 -0.554499600681715
-0.945040227246764 -0.540482879722995
-0.891078293715587 -0.527393121984623
-0.840476245453315 -0.515154431453848
-0.793026200995634 -0.503690475142763
-0.748532918074298 -0.49292486249818
-0.706812937888655 -0.482781546580713
-0.66769379222117 -0.473185241725975
-0.631013269643741 -0.464061852316497
-0.596618737227567 -0.455338907257268
-0.564366514322062 -0.446945994761242
-0.534121295109534 -0.438815192114016
};
\addplot [semithick, green01270, mark=*, mark size=3, mark options={solid}, only marks]
table {%
2.60254065376184 5.06573298882138
};
\addplot [semithick, darkturquoise23190207]
table {%
2.60254065376184 5.06573298882138
3.04712189375383 4.59312407763726
3.43444019898082 4.14607832703914
3.76826641302053 3.72305195732276
4.05396773463842 3.31714730232614
4.29011605836234 2.93964192723662
4.48443181136752 2.57752870202447
4.63409303855757 2.24544506082801
4.74726839565489 1.92919239982981
4.82145752736874 1.64226567057369
4.86440377027236 1.37123869379071
4.87727889709666 1.11840200967858
4.85996668864025 0.890397719899976
4.8171112033874 0.687421729891267
4.75248617228171 0.49396919919901
4.66341632898379 0.327662282569169
4.56239508153273 0.185718992276481
4.43793993524443 0.0568888647848566
4.29988131450554 -0.0573085828570153
4.14878997325561 -0.154095204863518
3.98675663467517 -0.234542395314576
3.81727817202891 -0.304295829122871
3.63980729385541 -0.358593020263111
3.45786956107799 -0.403553418387665
3.27297012428443 -0.44103454875922
3.08484412130565 -0.468181151806704
2.89327935124456 -0.481553754896288
2.70112044056324 -0.48592553159911
2.51353444219523 -0.487931470786067
2.33820016087937 -0.483280989748948
2.17471752622373 -0.48168298201873
2.02196440836566 -0.474251375601009
1.87977945554854 -0.475396996947278
1.74699184472942 -0.474778495356392
1.62292499092385 -0.469939620570905
1.5073180562255 -0.468796295816723
1.39941825657436 -0.467265477760571
1.2985648381125 -0.460548093086022
1.20469338472988 -0.458734859606579
1.11695334596794 -0.452206571055922
1.03510711708385 -0.443966442469494
0.958990006114917 -0.440292608157429
0.887993175680601 -0.436346681991437
0.821568865684888 -0.425531486876441
0.759858921401874 -0.418399361886422
0.702353665915441 -0.410660623660569
0.648866738078169 -0.404945590164159
0.598826457310871 -0.392752639492936
0.55261760170217 -0.390079336037774
0.509273811212847 -0.379412208073593
0.469253944252684 -0.377101724611883
};
\addplot [semithick, green01270, mark=*, mark size=3, mark options={solid}, only marks]
table {%
5.50926422486168 -1.85014478944775
};
\addplot [semithick, steelblue31119180]
table {%
5.50926422486168 -1.85014478944775
5.20490336752438 -1.84444152723631
4.9384980247796 -1.80573655392171
4.6043227333152 -1.78319939260677
4.28991082501829 -1.74926340277393
3.97548940004674 -1.72152930254895
3.6826314762983 -1.69935064860673
3.40944670137776 -1.6712917884478
3.15479447146308 -1.63975008136274
2.91766291875562 -1.60996177914531
2.69701279181098 -1.58679684929322
2.4913787496999 -1.56205295915251
2.29968102581095 -1.5324627304872
2.12112524474399 -1.49981255521525
1.95497272219637 -1.46686176186477
1.80056233802374 -1.4386252658028
1.65704449025469 -1.41533509270917
1.52328163578686 -1.38656488116197
1.3987983109576 -1.35504880339178
1.28335834608756 -1.33136962410165
1.17579888832297 -1.30189165600312
1.07583674826831 -1.27129853702192
0.983159558106896 -1.24509522682688
0.897166957349661 -1.22186175214433
0.817161252278081 -1.1951795384595
0.742881848432338 -1.16766012137103
0.67404930246069 -1.14207842543802
0.610050885144558 -1.11158233357938
0.55113554153753 -1.09158790620013
0.496347666569024 -1.06801243191675
0.445723323600555 -1.04880477921742
0.398527146091666 -1.02181039671923
0.354980110046627 -0.997217053738701
0.314752224448869 -0.973498754549869
0.277495628844495 -0.94712627869449
0.243363370687191 -0.927757940350697
0.211688365434246 -0.904517516954919
0.182501267302916 -0.88156117685718
0.155802710310392 -0.864232161259943
0.131086948229102 -0.844691723593916
0.108196633223694 -0.821052886577337
0.0873115639540161 -0.800707090346154
0.0680721422854679 -0.778220732670149
0.0505235525610321 -0.757372731495505
0.0344793239035896 -0.736667198597017
0.0198771320505762 -0.717359663478427
0.00655264416995734 -0.698008977106179
-0.00544336035770861 -0.682843111298653
-0.016486048614756 -0.664974315363195
-0.0265583910525151 -0.645579955032515
-0.0356030309801086 -0.6277179779299
};
\addplot [semithick, green01270, mark=*, mark size=3, mark options={solid}, only marks]
table {%
-0.809503978690769 5.06218697698483
};
\addplot [semithick, darkorange25512714]
table {%
-0.809503978690769 5.06218697698483
-0.35098815263033 4.73659865460906
0.0666080624737655 4.42454699388936
0.445001443801959 4.12663628703254
0.788413017666097 3.83607480846985
1.09745608950381 3.55484811718674
1.37408008695314 3.28143552642169
1.61669314462382 3.02436274257898
1.82919812290996 2.77805623110487
2.01387899029725 2.54012709306929
2.16867308595737 2.3203906669017
2.29766467467612 2.11297580545333
2.40575553158766 1.90785775837447
2.48664577913917 1.72553386291442
2.54637971650078 1.55531983477935
2.58970634130142 1.38733394128849
2.61151440690495 1.23826840959363
2.61808151207423 1.0954437374699
2.60733356051549 0.967467368822731
2.58340637926424 0.846766128856268
2.54563049525645 0.737914934630676
2.49461314078052 0.642889164011098
2.43447022750055 0.553688619453372
2.36289870443797 0.479157091343294
2.28465750784286 0.409494229162692
2.19735714435113 0.353421499707347
2.10294806359606 0.309092654686508
2.00450539903762 0.270735702457433
1.90226016570995 0.239405900614863
1.7983370538819 0.210777174185375
1.68814587353105 0.199293385267834
1.57917861547682 0.188974273598369
1.47721977060378 0.181373646277849
1.38206862969901 0.169517218802004
1.29286947106368 0.163870551544305
1.20978418337807 0.150053859043152
1.13193382978683 0.138762026835063
1.05900920475903 0.130011592435059
0.990842962537454 0.120035536943412
0.926836481123748 0.117032276546538
0.867057348482419 0.11368261366163
0.811166138519685 0.111622459505603
0.759083051136968 0.105888044413245
0.710395022636792 0.0996401592903298
0.664717047170898 0.0975907943812353
0.622168442994313 0.0918217469852401
0.582237100194019 0.0899822651305779
0.545061156664406 0.0842504486771802
0.510215945418851 0.080234619916022
0.477756752886482 0.0722022614755082
0.44730529436123 0.0650191837080452
};
\addplot [semithick, green01270, mark=*, mark size=3, mark options={solid}, only marks]
table {%
-3.95509449913879 -4.77936646051799
};
\addplot [semithick, forestgreen4416044]
table {%
-3.95509449913879 -4.77936646051799
-4.35799705743526 -4.28290241631555
-4.70225079137206 -3.81237267382945
-4.993455377118 -3.36096087978635
-5.23286510336824 -2.93274099363951
-5.42866456248239 -2.5399885828742
-5.59223454247219 -2.17303282596252
-5.74024950590641 -1.83722041072938
-5.86944610201857 -1.52816211872634
-5.96044228807307 -1.23573557911165
-6.05400110091771 -0.976540370127438
-6.09276245639465 -0.723015908760866
-6.15501538277646 -0.509306188161769
-6.15085318068369 -0.29261532141273
-6.12469061085547 -0.0952533262704864
-6.10773492613896 0.0713610642958233
-6.05084642621677 0.231557303731477
-5.98044808826666 0.373093700485061
-5.88646233961461 0.501620084900039
-5.7744010418361 0.615163617660903
-5.67392867662614 0.702604715009156
-5.55442626922066 0.780353043275072
-5.40094940576667 0.854529276504842
-5.24806850678315 0.909601714679491
-5.06475393049046 0.960577038795338
-4.87058520955914 0.998248064137466
-4.62301267705264 1.04082322706038
-4.3608657208714 1.07513445673694
-4.09310730686338 1.08691773085951
-3.82807309327245 1.0949441686907
-3.56194655824042 1.0870487089453
-3.307121549749 1.07792128019626
-3.06933230083268 1.0661443219236
-2.84759958615756 1.05527030762089
-2.64081716356434 1.04460261820149
-2.44781657712391 1.02884927421908
-2.26778077952168 1.00864720816071
-2.10028805924819 0.995609255103972
-1.94419501196826 0.983664239473202
-1.79854106619585 0.967064265054101
-1.66297558761066 0.954281207458311
-1.53646759104457 0.935985454601159
-1.4187909554991 0.921238786593957
-1.30926861215885 0.909026798685334
-1.20698897771196 0.889088340829325
-1.11190450266957 0.871190423351536
-1.0235693515457 0.857392420701961
-0.941194540258204 0.839116665253037
-0.864799177407262 0.827154988909276
-0.793558469191209 0.811304751856288
-0.72713088205477 0.789694764681456
};
\addplot [semithick, green01270, mark=*, mark size=3, mark options={solid}, only marks]
table {%
4.1849124674212 -3.90464512218038
};
\addplot [semithick, crimson2143940]
table {%
4.1849124674212 -3.90464512218038
3.84176457675885 -3.82924843386074
3.51206363256387 -3.74181292495718
3.19244571288934 -3.64384982745619
2.88954585572925 -3.55299163210423
2.59516977515631 -3.44758153858826
2.31298112636008 -3.33529690383831
2.04910091337295 -3.23328609723473
1.79588924963277 -3.12243693970765
1.5559613025403 -3.0088781041647
1.32833731969212 -2.88974128814214
1.11830840181858 -2.78004074168126
0.922332247418266 -2.67260897479267
0.736477126350702 -2.55709951482832
0.568012858708915 -2.45365968497597
0.410310310259887 -2.34659187286191
0.263835178806537 -2.23695688832428
0.131385885422774 -2.13305506658701
0.0129653524394688 -2.03713197342138
-0.0959023288497237 -1.93867233199888
-0.190927809972068 -1.85032937395542
-0.275223969570318 -1.76617862790716
-0.349916796887538 -1.68453026754656
-0.416410079987762 -1.60237520933871
-0.474239703289581 -1.5214212621059
-0.51931299058924 -1.45463757849344
-0.558356479540943 -1.3863136634625
-0.589137236623131 -1.32303925426471
-0.612902314155493 -1.26284529738878
-0.629012748599938 -1.2088067904682
-0.64033764180919 -1.15437386034345
-0.645675713028543 -1.10328769211186
-0.642518513043883 -1.06419883117671
-0.637914710555187 -1.01964864401733
-0.626589775671208 -0.984029036771585
-0.611864324254669 -0.950100176308596
-0.594917282482832 -0.91497156163437
-0.573308814652013 -0.885724029760945
-0.549939096711339 -0.861657110779513
-0.527886021942721 -0.833160461968794
-0.506707651208557 -0.809112911191322
-0.486328713993594 -0.792253302893403
-0.467059998063122 -0.774764276173345
-0.44894425531935 -0.752684080127704
-0.431760193363987 -0.728252915281331
-0.415052070650982 -0.712614404143264
-0.399431942164819 -0.69000074692273
-0.384456347600952 -0.66855181456006
-0.370029004391859 -0.650513523665681
-0.356345743535171 -0.630312393341335
-0.34301001737478 -0.617725207773312
};
\addplot [semithick, green01270, mark=*, mark size=3, mark options={solid}, only marks]
table {%
4.53989839301795 2.40937239619072
};
\addplot [semithick, mediumpurple148103189]
table {%
4.53989839301795 2.40937239619072
4.67100902496912 2.08665711021638
4.76449152086386 1.78658568928807
4.82309418932049 1.50882566634143
4.84951779826841 1.25294896446148
4.84640481561594 1.01844224382491
4.81632985655858 0.804716885933303
4.76383405367689 0.611548147106723
4.68970145832267 0.437749552874188
4.59355613182241 0.281982406313556
4.47979691465291 0.144069540208124
4.35064659191316 0.0232065895856111
4.20813592789273 -0.0814602475154383
4.05419060598264 -0.170801879072074
3.8906295308519 -0.245699783240611
3.71916387549226 -0.307039933067689
3.54139681936588 -0.355707180197624
3.35882392436394 -0.392580094055842
3.17283409596066 -0.418526251146333
2.98471107780864 -0.434397967382895
2.79563542905531 -0.441028464776074
2.60668693485008 -0.439228462317043
2.42590767704896 -0.434198076250107
2.25724151030683 -0.428864619317987
2.09988939770639 -0.423324789629347
1.95310371766073 -0.41766653240305
1.81618497868934 -0.411969072470719
1.68847875325053 -0.406302969304435
1.5693728158011 -0.400730198354766
1.45829447116432 -0.395304261936436
1.3547080601453 -0.390070332337337
1.25811263014201 -0.385065429255583
1.16803975926475 -0.380318633093546
1.08405152319942 -0.375851335061938
1.00573859473349 -0.371677524475882
0.932718466511381 -0.367804113063125
0.864633788199526 -0.364231295557144
0.80115080982363 -0.360952945319176
0.741957923593195 -0.357957043228079
0.686764297052956 -0.355226137599373
0.63529859089916 -0.35273783244924
0.587307755271898 -0.350465301009321
0.542555898784228 -0.348377821027396
0.500823224975615 -0.346441328060727
0.461905031282251 -0.344618982685598
0.425610766000864 -0.342871747310916
0.391763139086449 -0.341158968097448
0.36019728296844 -0.339438957348867
0.330759959894921 -0.3376695716572
0.303308812620804 -0.335808781054018
0.27771165554403 -0.333815224439709
};
\addplot [semithick, green01270, mark=*, mark size=3, mark options={solid}, only marks]
table {%
2.59246747582498 5.45904872508707
};
\addplot [semithick, sienna1408675]
table {%
2.59246747582498 5.45904872508707
3.08015057616621 4.95874152225369
3.50840627873917 4.48100251017
3.88006703132729 4.02643958705124
4.19802477596785 3.59548095122357
4.46521013067967 3.18838805988273
4.68457311892242 2.80526843252812
4.85906542309587 2.44608824417239
4.99162413242032 2.11068465854034
5.09262214426756 1.80034764657061
5.17457409030126 1.51623605136174
5.23799326423322 1.25632859324881
5.28321897877865 1.01876343354222
5.31041874349513 0.801829030986029
5.31958991184453 0.603955831486216
5.31056079819454 0.423708716095238
5.28299125656543 0.259780138277699
5.23637270409225 0.110983887312079
5.17002756340948 -0.0237505796737186
5.08310808947158 -0.145383289061376
4.97459453769163 -0.254768432204213
4.84329262170364 -0.352658618522135
4.68783020051964 -0.439708736711805
4.50665312634543 -0.516479455629877
4.30215757837104 -0.582570373531228
4.0911357658506 -0.634693964750152
3.87522511011874 -0.673809339680348
3.65594487382373 -0.700858737955514
3.43469846193517 -0.716763527654035
3.21277618769178 -0.72242069857601
2.99135844619199 -0.71869983051789
2.78046724228323 -0.712035836701521
2.58372592498592 -0.705222287596126
2.40019620583529 -0.698320525578607
2.22900099816815 -0.691381545500998
2.0693205209933 -0.684446519359344
1.92038865482305 -0.677547344006942
1.78148953230286 -0.670707211521301
1.6519543476077 -0.663941201272669
1.53115836963988 -0.657256892211842
1.41851814506756 -0.650654993400152
1.31348887819022 -0.64412999034855
1.21556197550941 -0.637670804318954
1.12426274372428 -0.631261461372436
1.03914823066329 -0.624881767628004
0.959805199409941 -0.618507986924826
0.885848226582642 -0.612113516861626
0.816917916389537 -0.605669559020751
0.752679222700378 -0.599145779072124
0.692819871960683 -0.592510952394135
0.637048880320482 -0.585733590844376
};
\addplot [semithick, green01270, mark=*, mark size=3, mark options={solid}, only marks]
table {%
5.17592935558338 -0.807092850715554
};
\addplot [semithick, orchid227119194]
table {%
5.17592935558338 -0.807092850715554
4.97628492621589 -0.87349933503892
4.75422631227554 -0.928281689503746
4.50871828980005 -0.971998329771227
4.25316102854626 -1.00207432553812
3.99505457752461 -1.01802636799459
3.73603116675101 -1.02097904119191
3.47758958333888 -1.01203636366777
3.22940034871852 -0.99746911861212
2.99794409141185 -0.981908554141043
2.78212593497886 -0.965481765123293
2.58092179543699 -0.948311829406561
2.39337372813955 -0.930517780501352
2.21858558354356 -0.912214532702126
2.05571895225893 -0.893512765361929
1.90398938097139 -0.874518773050881
1.76266284195002 -0.855334288301518
1.63105243989324 -0.836056283573675
1.50851534084079 -0.81677675895755
1.39444990878795 -0.797582521975413
1.28829303648783 -0.778554965640376
1.18951765772226 -0.759769850685451
1.0976304290661 -0.741297097589025
1.01216956986793 -0.723200593695637
0.932702849825478 -0.705538020365804
0.858825714150644 -0.688360704688216
0.790159536899759 -0.671713499855051
0.726349993592753 -0.655634697839891
0.667065544762871 -0.640155977531551
0.611996022569176 -0.6253023909702
0.560851313069491 -0.611092389808779
0.513360127193715 -0.597537893587554
0.469268853878623 -0.584644400867318
0.428340489226815 -0.572411143722214
0.39035363593607 -0.560831285551271
0.355101567612231 -0.549892161633367
0.322391352930013 -0.539575561328489
0.292043034942868 -0.529858050323439
0.263888861166049 -0.520711330837322
0.23777256036697 -0.512102637245629
0.213548662294522 -0.503995164155655
0.191081856864616 -0.49634852357441
0.170246389593176 -0.489119227456544
0.150925490330502 -0.48226119160752
0.133010832602465 -0.475726256648992
0.116402021104497 -0.469464721531691
0.101006105123942 -0.463425884907946
0.086737115884922 -0.457558589552774
0.0735156260175515 -0.451811764950409
0.0612683295499285 -0.446134963142551
0.0499276410068071 -0.440478882965684
};
\addplot [semithick, green01270, mark=*, mark size=3, mark options={solid}, only marks]
table {%
2.05991675650915 0.206539155446347
};
\addplot [semithick, gray127]
table {%
2.05991675650915 0.206539155446347
1.95153928317467 0.174058265532516
1.83776515025132 0.155395798215134
1.72138440531303 0.14606102712463
1.60920531635759 0.140526275664166
1.5046235938069 0.128357892754566
1.40679135917354 0.116951593921165
1.31510477466845 0.111507409386656
1.22965187314448 0.0994286906967299
1.14946529714832 0.0946873747089785
1.07455072638449 0.0897759297851064
1.00448723649914 0.0867900229456588
0.939255807782203 0.077174542697161
0.878079158755742 0.0718800157379443
0.821036806907613 0.0621172862621119
0.767333561586191 0.0619226620094016
0.717272844810916 0.0595996230377704
0.67072680404397 0.0505024626651233
0.627052640852181 0.0440490060177025
0.58623990824906 0.0358951838629079
0.547933328106894 0.0303461707681824
0.51216469039624 0.0224797110628341
0.478615204625749 0.0158277419432956
0.447025302867097 0.0144866539496772
0.417555145669465 0.0120011886758277
0.390053384807202 0.00812058890092879
0.364373730966776 0.00262677339565832
0.340174416714351 0.00161066398501472
0.31746907955753 0.00378362588545509
0.296545187699823 -0.0013731881705061
0.276742729463582 -0.00056893764068328
0.258253285960042 0.000827108138553285
0.241200842239492 -0.00350347856026576
0.224951927544925 0.000413070093991571
0.210058168007925 -0.00225450766096638
0.196040866629388 -0.00262703818343913
0.183080979648238 -0.00705001664117232
0.170875730675991 -0.0102679802707565
0.159254910298489 -0.0079327370314017
0.148669655785132 -0.0126134233502128
0.138683990093842 -0.0163151861127912
0.129137195944258 -0.0147758346102363
0.120292717909084 -0.0145471126280601
0.112207917338786 -0.0195358620315439
0.104300388586727 -0.0157015760980277
0.0969723914012667 -0.0116146857334264
0.0903894497729782 -0.0135328945572435
0.0842754738010783 -0.0173542710289273
0.078293313962162 -0.0144574562532905
0.0726910274439318 -0.00954062698296585
0.0678075446941321 -0.012991050055301
};
\addplot [semithick, green01270, mark=*, mark size=3, mark options={solid}, only marks]
table {%
-1.33900657467185 -1.99298496561491
};
\addplot [semithick, goldenrod18818934]
table {%
-1.33900657467185 -1.99298496561491
-1.4365810304162 -1.8418650573865
-1.51539472952826 -1.70273509939638
-1.57925756478746 -1.56834532476231
-1.62635479918541 -1.44613332205663
-1.65959448704768 -1.33147086174813
-1.67879128330238 -1.22759920476648
-1.68599936510086 -1.13192793968886
-1.68145669227984 -1.04650008323946
-1.66699551773602 -0.969119356473156
-1.64795233205217 -0.886498017230863
-1.61576176497226 -0.823054042663687
-1.57838733642339 -0.760690118214355
-1.53529460755753 -0.701559592538482
-1.48462335697821 -0.652339682703733
-1.42856056252161 -0.609461946416872
-1.36909752149637 -0.568911353048791
-1.30293200492914 -0.541298369767136
-1.23288211620371 -0.52189719689781
-1.16162101733673 -0.505054989784893
-1.09272172165262 -0.490317558256924
-1.028307940972 -0.471531633281882
-0.967875116673435 -0.453122272930115
-0.910964297661826 -0.441334777768751
-0.85764792387234 -0.429852619397026
-0.807693473373714 -0.418610607020725
-0.760998828451152 -0.403917917185399
-0.717262558014852 -0.386697770458008
-0.676034822337451 -0.373599945718743
-0.637302403253585 -0.361857455858632
-0.60101031637304 -0.348664110908066
-0.566993391833507 -0.333381493339026
-0.534913536874154 -0.320805697691056
-0.504565001128309 -0.314678238113704
-0.47631238396385 -0.303190993019011
-0.449630209763326 -0.295750222185846
-0.424660417321053 -0.286653767330969
-0.40106626187218 -0.282042990962226
-0.378984587992503 -0.276998716372988
-0.35823631524101 -0.27375499963137
-0.339073020801774 -0.262560911874366
-0.320744924666936 -0.259615655924855
-0.303752527545142 -0.251621498980048
-0.287721477660537 -0.244996985470237
-0.272539378826254 -0.241933612956189
-0.258353296495169 -0.237831052348903
-0.244984329950411 -0.235693080886416
-0.232551127130338 -0.231082210445673
-0.220886874570452 -0.226038486331319
-0.21010946205152 -0.21505905711182
-0.199825065222666 -0.205666409702968
};
\addplot [semithick, green01270, mark=*, mark size=3, mark options={solid}, only marks]
table {%
-2.50964808486869 1.22643416559638
};
\addplot [semithick, darkturquoise23190207]
table {%
-2.50964808486869 1.22643416559638
-2.32226067496943 1.20317590160868
-2.14792458313135 1.18614396221743
-1.98533475240473 1.16480585732885
-1.83393621492947 1.14414003684459
-1.69293059860551 1.12282770654819
-1.56171173969017 1.10324296058422
-1.43955803954256 1.08406546259703
-1.32582464716762 1.06421654732052
-1.21994360313234 1.04314348742268
-1.12144230845261 1.02181809774164
-1.02986177956054 1.00113561207693
-0.944866668434818 0.985345305467485
-0.865590400450366 0.9634949326938
-0.792143624090377 0.947950654348704
-0.723834946818196 0.932364864330672
-0.660096663334394 0.909668223907165
-0.600927477120963 0.885812845792414
-0.546279788274527 0.86825318733094
-0.495386427984404 0.84554238338905
-0.448229671509795 0.822391163529289
-0.404630061587617 0.800753125418573
-0.364530838343122 0.787013912584865
-0.327278544037871 0.771950605318037
-0.292508600811823 0.749436551220874
-0.260556816633626 0.731446725899218
-0.230905808773197 0.709917873701747
-0.203858717309166 0.697365890349388
-0.178771737212897 0.683537112924216
-0.155548392776429 0.668864423109018
-0.134228551698404 0.658057859288203
-0.11424804354965 0.639359363990952
-0.0960678257264625 0.626057387363413
-0.0793712626386475 0.614678362835594
-0.0638343089653022 0.599134172038069
-0.0496541337548788 0.585939306389639
-0.0364617015005528 0.56754607290823
-0.0244955002254637 0.550890384631762
-0.0135938185664569 0.534331979810378
-0.00375359343742841 0.520040807682745
0.0051688710973841 0.506859484953981
0.0133056956029689 0.493119060987089
0.0205765313585963 0.482691148506993
0.0272743400450189 0.46986233149164
0.0332618523492493 0.458927460561825
0.0387903461702697 0.44444845631418
0.0436537996597605 0.432065355359939
0.0478258942416073 0.425026250017575
0.0517811942080909 0.412583216175741
0.0553628174131355 0.397412488849958
0.058331604694818 0.38641755930155
};
\addplot [semithick, green01270, mark=*, mark size=3, mark options={solid}, only marks]
table {%
2.22801544516873 -1.91135733917186
};
\addplot [semithick, steelblue31119180]
table {%
2.22801544516873 -1.91135733917186
2.04804303627629 -1.87310046906282
1.88065783924826 -1.83456258929644
1.7250277553717 -1.79580514902918
1.58037515327735 -1.7568899214545
1.44597323865145 -1.71787910711914
1.32114266201363 -1.67883537052187
1.20524834989121 -1.6398218140916
1.09769654564277 -1.60090189396717
0.997932047039348 -1.56213928230231
0.905435628503884 -1.52359768109119
0.819721636644231 -1.48534059274992
0.740335748395427 -1.4474310528926
0.666852881717689 -1.40993133090472
0.598875249381321 -1.37290260403917
0.536030546912224 -1.33640461083809
0.477970266275249 -1.30049528971689
0.424368127340739 -1.26523040853197
0.374918619615148 -1.23066319089268
0.329335647122665 -1.19684394486849
0.28735126970395 -1.16381969958652
0.24871453435305 -1.13163385501205
0.213190390546629 -1.10032584995782
0.180558683833064 -1.06993085307824
0.150613222244865 -1.04047948127446
0.123160910378021 -1.0119975495687
0.0980209462481695 -0.984505856104564
0.0750240762873774 -0.958020005497139
0.0540119040883575 -0.9325502732977
0.0348362487363064 -0.908101513855851
0.0173585487934757 -0.884673113362089
0.00144930821901075 -0.862258989340564
-0.013012419282594 -0.840847637340039
-0.0261395107861382 -0.820422225045746
-0.0380372226455112 -0.800960733510855
-0.0488036098758166 -0.782436144688793
-0.058529919134529 -0.764816673941437
-0.067300957832059 -0.74806604570825
-0.0751954417197392 -0.732143810052328
-0.0822863231247803 -0.717005697355666
-0.0886411018271776 -0.702604008021916
-0.0943221204029702 -0.688888033664574
-0.0993868456918073 -0.675804505915403
-0.103888137884679 -0.663298068685472
-0.107874508570182 -0.651311769452089
-0.111390368925107 -0.639787564931908
-0.114476269087821 -0.628666836335292
-0.117169129611183 -0.617890909281437
-0.119502465756029 -0.607401573388665
-0.121506605256892 -0.597141596540349
-0.123208900069041 -0.587055228864114
};
\addplot [semithick, green01270, mark=*, mark size=3, mark options={solid}, only marks]
table {%
2.03053050095627 0.819732303586688
};
\addplot [semithick, darkorange25512714]
table {%
2.03053050095627 0.819732303586688
1.99077889523384 0.740575912032964
1.94194166327199 0.670537078530331
1.88510087953314 0.609218275624348
1.82129073944473 0.556203316583392
1.75149558953592 0.511061130667184
1.67664837629655 0.473349361609164
1.59762949242117 0.442617784616887
1.51526599885688 0.418411537910583
1.43033120090452 0.400274165509956
1.34354455652226 0.387750468648062
1.26052815219413 0.377289941331172
1.18284923017639 0.367669210909934
1.11016834823886 0.358800698925686
1.04216724318885 0.350598205361927
0.978547430231359 0.342977272178478
0.919028900993918 0.33585555965245
0.863348914027692 0.329153229860331
0.811260871917201 0.322793331686337
0.762533279429701 0.316702181841197
0.716948777413976 0.310809736522342
0.674303247419409 0.305049948539937
0.634404982251405 0.299361104972044
0.59707391791066 0.293686140694196
0.562140922582789 0.287972923451663
0.529447138553308 0.282174506503573
0.498843373122058 0.276249345263754
0.470189534782348 0.270161474789927
0.443354111114414 0.263880645426788
0.418213685021303 0.257382414385327
0.394652486108913 0.250648191535887
0.372561974181361 0.243665238201332
0.351840451988814 0.236426618254352
0.332392704527925 0.228931101344481
0.314129662355452 0.221183018600763
0.296968086533847 0.213192071670253
0.280830272983807 0.204973096455608
0.265643774173025 0.19654578340215
0.251341136222805 0.187934356651184
0.237859649664618 0.179167214817534
0.22514111222705 0.170276536560967
0.213131602179686 0.161297854499372
0.201781260903993 0.152269601352633
0.191044083501979 0.143232632506761
0.180877716390837 0.134229729445131
0.171243260965657 0.125305088705118
0.162105082542128 0.116503801182063
0.153430623916463 0.107871326716716
0.145190223000254 0.0994529689660671
0.137356934102966 0.0912933555702367
0.129906352544039 0.0834359285897586
};
\addplot [semithick, green01270, mark=*, mark size=3, mark options={solid}, only marks]
table {%
0.217118631662959 2.18368891186708
};
\addplot [semithick, forestgreen4416044]
table {%
0.217118631662959 2.18368891186708
0.340853908758262 2.06547925097778
0.450347709070814 1.95272586387531
0.546285919594694 1.84556546947569
0.629369247363937 1.74408737430197
0.700307699023162 1.64833702266908
0.759815468526986 1.55831954145485
0.808606231247336 1.47400326054759
0.847388840817172 1.39532319115271
0.876863423213145 1.32218444524773
0.897717860877706 1.25446558059653
0.910624658102263 1.19202185686751
0.916238177436601 1.13468838954642
0.91519223555493 1.08228318949038
0.908098045795116 1.03461007713424
0.895542493493805 0.991461461530345
0.878086729265555 0.952620975576927
0.856265064517718 0.91786595996518
0.830584152753541 0.886969789547188
0.801522439592622 0.859704036993783
0.76952986392899 0.83584046976849
0.73502779225108 0.815152877587526
0.69958618642772 0.796682721176945
0.666202614620537 0.77852324057216
0.634746742601118 0.76061700331226
0.605095842270398 0.742911740549683
0.577134359605818 0.725360574697791
0.550753516285308 0.707922206961527
0.525850942192125 0.690561062125914
0.502330336134056 0.673247388449245
0.480101152238306 0.655957310998704
0.459078309609973 0.638672837270856
0.439181922967899 0.621381814453386
0.420337052097221 0.604077838202414
0.402473468083351 0.586760113326789
0.385525434417547 0.569433267281881
0.36943150118981 0.552107117875632
0.35413431071045 0.534796397073982
0.339580413027284 0.517520433256713
0.325720089930786 0.500302794713573
0.312507186164508 0.483170897580025
0.299898946682215 0.466155581788155
0.287855858916265 0.449290658947389
0.276341499143311 0.432612436368527
0.265322382152926 0.416159221700213
0.254767813541893 0.39997081285685
0.244649744071061 0.384087978079148
0.234942625632377 0.36855193108135
0.225623268480448 0.353403806301695
0.216670699485227 0.338684139284245
0.208066021259675 0.324432357180749
};
\addplot [semithick, green01270, mark=*, mark size=3, mark options={solid}, only marks]
table {%
0 0
};
\addplot [semithick, crimson2143940]
table {%
0 0
0 0
-1.64879440014968e-05 0.000515566591351217
-3.96174727482888e-05 0.00152893569860268
-6.03296409789943e-05 0.00301764151770878
-7.03821307619e-05 0.00495473764488464
-6.23744369852586e-05 0.00730906281578789
-2.9768501276642e-05 0.0100455486199537
3.30956661409344e-05 0.0131255661344449
0.000130989770467851 0.0165073080303569
0.000267792746497684 0.0201462023500323
0.000446492255447724 0.0239953538351363
0.000669192343614742 0.0280060084092955
0.000937127367767823 0.0321280361865611
0.00125068221771451 0.0363104281908392
0.00160941879246856 0.040501801833498
0.00201210861359675 0.0446509101079751
0.00245677138830908 0.0487071494222602
0.00294071926634771 0.0526210610029816
0.00346060646936791 0.0563448208683573
0.00401248390990447 0.0598327134808368
0.00459185835975855 0.0630415843527258
0.00519375567526217 0.0659312670878203
0.00581278753987174 0.0684649805969878
0.00644322114334857 0.0706096925231643
0.00707905118178858 0.0723364452484168
0.00771407353428726 0.0736206412291776
0.00834195995033256 0.0744422848117563
0.0089563330672995 0.0747861781147171
0.00955084106980452 0.0746420690233363
0.010119231302219 0.0740047498195317
0.0106554221523287 0.0728741054636164
0.0111535725378726 0.0712551110470342
0.0116081483483499 0.069157778442869
0.012013985221825 0.0665970526882995
0.0123663470701979 0.0635926591352378
0.0126609798061993 0.0601689028971129
0.0128941597707923 0.056354422596235
0.0130627364102668 0.052181900872638
0.0131641688075681 0.0476877345471788
0.0131965557317543 0.0429116677346678
0.0131586589323239 0.0378963915728789
0.013049919470854 0.0326871145667407
0.0128704669502844 0.0273311078405208
0.0126211215715792 0.0218772298414355
0.0123033890177023 0.0163754352433526
0.0119194482351399 0.0108762729570502
0.0114721322528933 0.00543037826227393
0.0109649022472377 8.79641355090053e-05
0.0104018151269105 -0.00510168314463306
0.00978748497709764 -0.0100906990752989
};
\end{axis}

\end{tikzpicture}

%% file: fig/comparison/outputs.tex
\begin{tikzpicture}

\definecolor{darkgray176}{RGB}{176,176,176}
\definecolor{darkorange25512714}{RGB}{255,127,14}
\definecolor{forestgreen4416044}{RGB}{44,160,44}
\definecolor{lightgray204}{RGB}{204,204,204}
\definecolor{steelblue31119180}{RGB}{31,119,180}

\begin{groupplot}[group style={group size=2 by 1}]
\nextgroupplot[
  width=0.49\textwidth,
  height=3.6cm,
legend cell align={left},
legend style={
  fill opacity=0.8,
  draw opacity=1,
  text opacity=1,
  at={(0.03,0.97)},
  anchor=north west,
  draw=lightgray204
},
tick align=outside,
tick pos=left,
x grid style={darkgray176},
xmajorgrids,
xmin=-2.45, xmax=51.45,
xtick style={color=black},
y grid style={darkgray176},
xlabel={Time step $k$},
ylabel={$\ee$},
ymajorgrids,
ymin=2.33554311802885, ymax=36.8744773892027,
ytick style={color=black}
]
\addplot [thick, black,  dashed]
table {%
0 4.37444990644978
1 4.88128604124862
2 5.32573983995805
3 5.75728365548636
4 6.17335637215803
5 6.56128183557693
6 6.9247452254787
7 7.26714629893458
8 7.59162654649103
9 7.90109449578038
10 8.19824940485442
11 8.4856035665944
12 8.76550342645672
13 9.04014969840757
14 9.31161664807959
15 9.58187069786377
16 9.85278849575347
17 10.1261745782247
18 10.4037787472126
19 10.6873132722861
20 10.9784700213984
21 11.2789376170645
22 11.5904187094801
23 11.9146474539205
24 12.2534072767405
25 12.6085490124323
26 12.9820094934808
27 13.3758306751917
28 13.7921793792661
29 14.233367742654
30 14.7018744621695
31 15.2003669304873
32 15.7317243655057
33 16.299062042655
34 16.9057567486017
35 17.5554735849585
36 18.2521942621118
37 19.0002470361411
38 19.8043384560966
39 20.6695871046533
40 21.6015595324418
41 22.6063086052302
42 23.6904145036683
43 24.8610286375921
44 26.1259207610128
45 27.493529599985
46 28.9730173336823
47 30.5743282993339
48 32.3082523243347
49 34.1864931240078
};
\addlegendentry{Ground Truth}
\addplot [semithick, steelblue31119180, opacity=0.8]
table {%
0 4.1867496959345
1 4.74071411767217
2 5.23059971897731
3 5.67062502905497
4 6.07981453098249
5 6.4637325236198
6 6.82558462234142
7 7.16835946228339
8 7.49484770759422
9 7.8076600839582
10 8.10924459369741
11 8.40190306100301
12 8.68780714397727
13 8.96901394015905
14 9.24748130303995
15 9.52508297874549
16 9.80362366455125
17 10.0848540842293
18 10.3704861693835
19 10.6622084309427
20 10.9617016008582
21 11.2706546208083
22 11.5907810523814
23 11.9238359818052
24 12.2716334918485
25 12.6360647740714
26 13.0191169561755
27 13.4228927218383
28 13.8496308041506
29 14.3017274386436
30 14.7817588679439
31 15.2925049973669
32 15.8369743093081
33 16.4184301541587
34 17.0404185467178
35 17.7067976097575
36 18.421768820572
37 19.1899102320898
38 20.0162118575131
39 20.9061134265525
40 21.8655447422446
41 22.9009688901627
42 24.0194285766695
43 25.2285958998361
44 26.5368258858907
45 27.9532141557106
46 29.4876591200937
47 31.1509291395214
48 32.9547351240423
49 34.9118090919964
};
\addlegendentry{\MGenSec{}}
\addplot [thick, darkorange25512714, ]
table {%
0 5.09485652393448
1 5.59391666161976
2 6.03827877160303
3 6.43136387209972
4 6.77638694100975
5 7.07636828155601
6 7.33414439050855
7 7.55237835512547
8 7.93312137524938
9 8.31685930766785
10 8.67681884004352
11 9.01359762124144
12 9.32776776408732
13 9.61987694170087
14 9.89044957971773
15 10.139988135448
16 10.3689744540953
17 10.5778711913336
18 10.767123290824
19 10.9371595046479
20 11.0883939441556
21 11.221227648372
22 11.3360501568743
23 11.4332410739663
24 11.513171611007
25 11.5762060939269
26 11.6227034232624
27 11.6530184744703
28 11.6675034268344
29 11.6665090099444
30 11.650385657512
31 11.6194845591657
32 11.5741586018501
33 11.5147631935121
34 11.4416569628959
35 11.3552023304659
36 11.2557659467239
37 11.1437189954775
38 11.0194373609251
39 10.8833016587485
40 10.7356971327279
41 10.5770134197001
42 10.4076441869638
43 10.2279866474756
44 10.0384409593668
45 9.83940951743723
46 9.63129614532557
47 9.41450519802085
48 9.18944058524079
49 8.95650472696587
};
\addlegendentry{\MStdSec}
\addplot [thick, forestgreen4416044, ]
table {%
0 3.90549467580948
1 4.47200575891274
2 4.96921065295443
3 5.42312975001541
4 5.84978306198083
5 6.25323003854101
6 6.63630420431361
7 7.0016677595983
8 7.35182574053205
9 7.68913968951123
10 8.01584094266613
11 8.33404363383739
12 8.6457575076373
13 8.95290062778998
14 9.25731206104726
15 9.56076461158838
16 9.86497767595178
17 10.171630284239
18 10.4823743895996
19 10.7988484648816
20 11.1226914628361
21 11.4555571944438
22 11.7991291787951
23 12.1551360175587
24 12.525367347437
25 12.9116904251658
26 13.3160674016135
27 13.7405733443975
28 14.1874150722016
29 14.658950868694
30 15.1577111496297
31 15.6864201634329
32 16.248018813319
33 16.8456886978799
34 17.4828774770655
35 18.1633256816845
36 18.8910950969816
37 19.6705988645631
38 20.5066334620028
39 21.404412735925
40 22.3696041822905
41 23.4083676870881
42 24.5273969617245
43 25.7339639302069
44 27.0359663498195
45 28.4419789735128
46 29.961308590772
47 31.6040533144444
48 33.3811665140292
49 35.3045258314221
};
\addlegendentry{\MLtiRnn{}}

\nextgroupplot[
  width=0.49\textwidth,
  height=3.6cm,
tick align=outside,
tick pos=left,
x grid style={darkgray176},
xlabel={Time step $k$},
xmajorgrids,
xmin=-2.45, xmax=51.45,
xtick style={color=black},
y grid style={darkgray176},
ymajorgrids,
ymin=-0.275983600812416, ymax=7.12674587722218,
ytick style={color=black}
]
\addplot [thick, black,  dashed]
table {%
0 2.60254065376184
1 3.08237331678773
2 3.51018154409206
3 3.88822045738023
4 4.21837973118774
5 4.50281478827423
6 4.74362352509099
7 4.94303921845839
8 5.10333431179883
9 5.22679937514028
10 5.31574277710108
11 5.37197934575113
12 5.39970887109831
13 5.39657194373681
14 5.36869869355959
15 5.33541453831052
16 5.25177375967827
17 5.16042373987777
18 5.05165981188622
19 4.92734527885726
20 4.78943082597908
21 4.63954732758655
22 4.47929493417937
23 4.31020439549083
24 4.133428129077
25 3.95017115389903
26 3.76193881751627
27 3.57016081870205
28 3.37579470367101
29 3.18011217986788
30 2.98428006713669
31 2.7893527622023
32 2.59596246656776
33 2.40446417289292
34 2.21592455920523
35 2.03099489850748
36 1.8500457169874
37 1.67392657348745
38 1.50308438575475
39 1.33789746226987
40 1.1790734247032
41 1.02669370632037
42 0.880724306708479
43 0.741779371380309
44 0.610156625692312
45 0.485742463240058
46 0.368488047378403
47 0.258670908259012
48 0.156055728070945
49 0.0605041027346113
};
\addplot [thick, steelblue31119180, ]
table {%
0 2.49312068027891
1 2.9941898310423
2 3.44781333851779
3 3.86586795254739
4 4.21160166642239
5 4.53771522265575
6 4.78755653507142
7 5.02392527375603
8 5.19016120061298
9 5.34519621241387
10 5.45681888221014
11 5.51849918983968
12 5.54805553586013
13 5.57662109130983
14 5.53877685681528
15 5.44768756044856
16 5.34634610983761
17 5.22615243080004
18 5.08545337019326
19 4.93462678288313
20 4.78418138132046
21 4.61319925450671
22 4.44644678000964
23 4.27489079266413
24 4.08983784084647
25 3.89243384837824
26 3.70159851278615
27 3.51737597668825
28 3.30689309445537
29 3.12368043158005
30 2.91765498843393
31 2.75171568730029
32 2.55771274227221
33 2.36118155094124
34 2.18969045505229
35 2.00992719631011
36 1.82278660624671
37 1.66803972825841
38 1.4920923058274
39 1.33157278107103
40 1.19328377272108
41 1.04657389138138
42 0.889884989855637
43 0.76930917418233
44 0.642745109909355
45 0.529503590228353
46 0.40081210640265
47 0.321607705747432
48 0.200664604035053
49 0.130679772697324
};
\addplot [thick, darkorange25512714, ]
table {%
0 2.60491773096259
1 2.8869320689325
2 3.22835425741454
3 3.55562023546003
4 3.81885795661704
5 4.06943543361737
6 4.25300850831206
7 4.43272259484086
8 4.55221821214227
9 4.81356419625584
10 5.09161857763866
11 5.2883773206467
12 5.46724278263789
13 5.74315666258605
14 5.8470445305921
15 5.97295550951072
16 6.13897688529212
17 6.28516701537568
18 6.38385848802216
19 6.47047994687435
20 6.58148110714571
21 6.62857964338046
22 6.70554256199289
23 6.77226863504212
24 6.79025817367515
25 6.75985254763465
26 6.75218271790272
27 6.76404242661123
28 6.66117948651981
29 6.65329814079707
30 6.53804875982697
31 6.56347228805833
32 6.45673223404766
33 6.31715424348635
34 6.25239453703967
35 6.12850534235007
36 5.94729110204915
37 5.86562951924738
38 5.66921006522378
39 5.50248733296402
40 5.39192825874625
41 5.21550330764458
42 4.96623737261381
43 4.82692110269989
44 4.6301261190841
45 4.45303426337735
46 4.1810326333057
47 4.07300713081713
48 3.76652041995402
49 3.63108920067478
};
\addplot [thick, forestgreen4416044, ]
table {%
0 2.31569140749541
1 2.8397927189691
2 3.30240512926389
3 3.70015905243375
4 4.06375340892163
5 4.35764983667952
6 4.62836183285717
7 4.83244576118711
8 5.01704879656643
9 5.14176285454432
10 5.23697880008864
11 5.30952770250401
12 5.35078652841184
13 5.34125010546786
14 5.33399738776294
15 5.29995737240166
16 5.22825371954842
17 5.13805049369087
18 5.03662502260072
19 4.91903176384827
20 4.78124522651935
21 4.64143985134477
22 4.48381279178306
23 4.31827407706172
24 4.15290341277823
25 3.98793660418775
26 3.81166220726861
27 3.60514843335876
28 3.32648205104686
29 3.08266210207232
30 2.84153839260984
31 2.63932423729701
32 2.43296230901817
33 2.2374848929643
34 2.06796282572324
35 1.90276702648755
36 1.74197934839387
37 1.60835667050137
38 1.46844185629742
39 1.34406791609958
40 1.23794635952104
41 1.13071106044619
42 1.02157111546404
43 0.937611727764418
44 0.852337715986515
45 0.777132564211194
46 0.694520224839313
47 0.643003471159609
48 0.567021887327652
49 0.522613135696294
};

\end{groupplot}

\draw ({$(current bounding box.south west)!0.5!(current bounding box.south east)$}|-{$(current bounding box.south west)!1!(current bounding box.north west)$}) node[
  scale=0.7,
  anchor=north,
  text=black,
  rotate=0.0
]{};
\end{tikzpicture}

%% file: content/conclusion.tex
In this work, we have presented a method for learning regionally stable nonlinear models from input-output measurements. We presented convex constraints that enable analyzing the regional stability of dynamic \glspl{rnn}. In addition, we proposed a learning algorithm that ensures these constraints are met during training by using barrier functions. Numerical experiments verify that the regionally stable models extend the class of learnable systems compared to globally stable dynamic \glspl{rnn}.

The derived parametric constraints are conservative, and future work will address this. Furthermore, we aim to use our identification framework to learn dynamics from real-world datasets.